\documentclass{amsart}
\usepackage[utf8]{inputenc}
\usepackage{amsmath}
\usepackage{array, amsfonts}
\usepackage{amsthm}
\usepackage{amssymb}
\usepackage{enumerate}
\usepackage{lineno}
\usepackage[bookmarks=true]{hyperref}
\usepackage[foot]{amsaddr}
\usepackage{enumitem}
\usepackage{mathrsfs}
\usepackage{stackengine}
\usepackage{bookmark}
\usepackage{amsrefs}
\usepackage{color,soul}
\usepackage[draft]{todonotes}
\usepackage{fancyhdr}
\usepackage{MnSymbol}
\usepackage{algorithm2e}
\usepackage[noend]{algpseudocode}
\usepackage{mathtools}

\definecolor{darkgreen}{rgb}{0,0.5,0}

\DeclareMathOperator*{\myDelta}{\Delta}

\newcommand{\comment}[1]

\newcommand\myfunc[5]{%
	\begingroup
	\setlength\arraycolsep{0pt}
	#1\colon\begin{array}[t]{c >{{}}c<{{}} c}
		#2 & \to & #3 \\ #4 & \mapsto & #5 
	\end{array}%
	\endgroup}

\title{ \Large Learning the undecidable \\ from networked systems}

\author{Felipe S. Abrah\~{a}o}
\address[Felipe S. Abrah\~{a}o, Klaus Wehmuth, Artur Ziviani]{National Laboratory for Scientific Computing (LNCC) \\ 25651-075 – Petropolis, RJ -- Brazil }
\email{fsa@lncc.br}

\author{\'{I}tala M. Loffredo D'Ottaviano}
\address[Ítala M. L. D'Ottaviano]{
	Centre for Logic, Epistemology and the History of Science, 
University of Campinas (UNICAMP) -- Brazil
}
\email{itala@cle.unicamp.br}

\author{Klaus Wehmuth}
\email{klaus@lncc.br}

\author{Francisco Antônio Dória} 
\address[Francisco Antônio Dória]{ Advanced Studies Research Group and Fuzzy Sets Laboratory, PIT, Production Engineering Program, COPPE, UFRJ \\ P.O. Box 68507, 21945-972  -- Rio de Janeiro, RJ -- Brazil}

\author{Artur Ziviani}
\email{ziviani@lncc.br}

\thanks{ Authors acknowledge the partial support from CNPq through their individual grants: F. S. Abrah\~{a}o (313.043/2016-7), K. Wehmuth (312599/2016-1), and A. Ziviani~(308.729/2015-3). Authors also acknowledge the INCT in Data Science -- INCT-CiD (CNPq 465.560/2014-8). Authors also acknowledge the partial support from CAPES, FAPESP, and FAPERJ}

\fancypagestyle{style1}{
\fancyhead{}
\fancyhead[L]{ Learning the undecidable from networked systems }
\fancyhead[R]{\thepage}
\fancyfoot{}
}
\fancypagestyle{plain}{
	\fancyhead{}
	\fancyhead[C]{ \textit{Research Report} \\ National Laboratory for Scientific Computing (LNCC)  \\ \textbf{Extended version of the paper:} }
}
\begin{document}

\maketitle%\thispagestyle{plain}	

\begin{abstract}\label{abstract}
	This article presents a theoretical investigation of computation beyond the Turing barrier from emergent behavior in distributed systems.  
	In particular, we present an algorithmic network that is a mathematical model of a networked population of randomly generated computable systems with a fixed communication protocol. 
	Then, in order to solve an undecidable problem, we study how nodes (i.e., Turing machines or computable systems) can harness the power of the metabiological selection and the power of information sharing (i.e., communication) through the network. 
	Formally, we show that there is a pervasive network topological condition, in particular, the small-diameter phenomenon, that ensures that every node becomes capable of solving the halting problem for every program with a length upper bounded by a logarithmic order of the population size. 
	In addition, we show that this result implies the existence of a central node capable of emergently solving the halting problem in the minimum number of communication rounds.
	Furthermore, we introduce an algorithmic-informational measure of synergy for networked computable systems, which we call local algorithmic synergy.
	Then, we show that such algorithmic network can produce an arbitrarily large value of expected local algorithmic synergy.
\end{abstract}



	% Appendix conditions

	\newtheorem{thmundersubsection}{Theorem}[subsection]
	\newtheorem{amslemmaundersubsection}{Lemma}[subsection]
	\newtheorem{amscorollaryunderthmundersubsection}{Corollary}[thmundersubsection]
	\newtheorem{corollaryundersubsection}{Corollary}[subsection]
	\newtheorem{amslemmaundersection}{Lemma}[section]

	\newtheorem{notation}{Notation}[subsection]

	\theoremstyle{definition}
	\newtheorem{amsproposition}{Proposition}[subsection]
	\newtheorem{amsdefinition}{Definition}[subsection]
	\newtheorem{amssubdefinition}{Definition}[amsdefinition]
	\newtheorem{subsubdefinition}{Definition}[amssubdefinition]
	\newtheorem{autodefinition}{Definition}
	\newtheorem{amsdefinitionnotation}{Notation}[amsdefinition]
	\newtheorem{amsnotationundersubsection}{Notation}[subsection]
	\newtheorem{amsnotationunderamsdefinition}{Notation}[amsdefinition]

	\theoremstyle{remark}
	\newtheorem{subnotation}{Notation}[amsdefinition]
	
	\theoremstyle{remark}
	\newtheorem{remark}{Remark}[amsdefinition]
	\newtheorem{noteunderamsdefinition}{Note}[amsdefinition]
	\newtheorem{noteunderamslemmaundersubsection}{Note}[amslemmaundersubsection]
	\newtheorem{noteunderthmundersubsection}{Note}[thmundersubsection]
	\newtheorem{subremarknote}{Note}[amssubdefinition]
	\newtheorem{subsubremarknote}{Note}[subsubdefinition]
	\newtheorem{subremarknote2}{Note}[noteunderamsdefinition]
	\newtheorem{note}{\textbf{Note}}[subsection]
	\newtheorem{noteunderamsnotationundersubsection}{Note}[amsnotationundersubsection]

	\theoremstyle{thm}
	\newtheorem{thmundersection}{Theorem}[section]

	\newtheorem{amscorollaryunderthmundersection}{Corollary}[thmundersection]
	\newtheorem{Bcorollaryundersection}{Corollary}[section]
	\newtheorem{amscorollaryunderamslemmaundersection}{Corollary}[amslemmaundersection]

	\theoremstyle{definition}
	\newtheorem{Bproposition}{Proposition}[section]
	\newtheorem{amsdefinitionundersection}{Definition}[section]
	\newtheorem{amsdefinitionunderamsdefinitionundersection}{Definition}[amsdefinitionundersection]
	\newtheorem{Bsubsubdefinition}{Definition}[amsdefinitionunderamsdefinitionundersection]
	\newtheorem{Bautodefinition}{Definition}
	\newtheorem{Bnotation}{Notation}[section]
	\newtheorem{BnotationunderBnotation}{Notation}[Bnotation]
	
	\theoremstyle{remark}
	\newtheorem{Bsubnotation}{Notation}[amsdefinitionundersection]
	\newtheorem{noteunderamsdefinitionundersection}{Note}[amsdefinitionundersection] 
	
	\theoremstyle{remark}
	\newtheorem{Bremark}{Remark}[amsdefinitionundersection]
	\newtheorem{Bremarknote}{Note}[amsdefinitionundersection]
	\newtheorem{Bnoteunderamslemmaundersection}{Note}[amslemmaundersection]
	\newtheorem{Bnoteunderthmundersection}{Note}[thmundersection]
	\newtheorem{Bsubremarknote}{Note}[amsdefinitionunderamsdefinitionundersection]
	\newtheorem{Bsubsubremarknote}{Note}[Bsubsubdefinition]
	\newtheorem{Bsubremarknote2}{Note}[Bremarknote]
	\newtheorem{Bnote}{\textbf{Note}}[section]
	\newtheorem{BnoteunderBnotation}{Note}[Bnotation]

\newpage

\pagestyle{style1}
	
\section{Introduction}\label{sectionIntro}

We study the general problem of how computable systems may take advantage of an uncomputable environment to solve undecidable problems, in particular, the halting problem.
Indeed, although the theory of Turing degrees is a well-established field in computability theory~\cite{Syropoulos2008,Crnkovic2012a,Calude2002} and mathematical logic~\cite{Rogers1987}, the possibility of computing beyond the Turing limit, that is, solving problems of Turing degree $\mathbf{0'}$ or above, is one of the major conundrums in the interface of theoretical computer science, mathematics, physics and biology.
Such subject has its roots on the incompleteness results in foundational mathematics, mathematical logic, or recursion theory.
This posits questions on the computability of the Universe~\cite{DaCosta2006,DaCosta2009a,Cooper2009,Longo2012,Prokopenko2019,Copeland2002,Crnkovic2012a,Calude2002}, living systems\cite{Longo2012,Prokopenko2019,Copeland2002,Abrahao2015,Shimansky2018}, or the human mind~\cite{Siegelmann1995a,Zenil2006a,Copeland2002,Copeland1998}.

On the other hand, besides artificial systems, the computable nature of Life has been supported by current findings and scientific views in computer simulation and evolutionary computation~\cite{Bedau1998,Standish2003,Fouks1999,Dingle2018} and algorithmic-informational theoretical biology~\cite{Chaitin2012,Chaitin2014,Hernandez-Orozco2018,Hernandez-Orozco2018a,Abrahao2015,Abrahao2016,Chaitin2013,Chaitin2018}.
In this way, as a pure mathematical work inspired by important concepts from complex systems science, the present article has an underlying objective of reconciling an hypothetical computable nature of Life (or artificial systems) with an hypothetical uncomputable Nature (or environment).

We study models of computation for synergistically solving an uncomputable problem in a network of computable systems.
We show an abstract model for networked computable systems that can harness the power of random generation of individuals and the power of selection made by an irreducibly more powerful environment.
In particular, we investigate the problem of how networked populations of randomly generated programs under metabiological selection of the fit in pervasive topological conditions can solve more and more instances of the halting problem by a fixed global communication protocol.
For this purpose, we investigate a class of algorithmic networks based on \cite{Abrahao2017published,Abrahao2018publishedAMS} that can solve the halting problem by a synergistic communication protocol of imitation of the fittest neighbor.
Thus, our theoretical results show that, within a hypercomputable environment as our assumption, whole populations of computable systems can be hypercomputers. 
To this end, we present definitions, theoretical models, and theorems based on computability theory, algorithmic information theory, distributed computing, and complex networks theory.

In this sense, this article shows how metabiology's early findings in \cite{Chaitin2012,Chaitin2013} have ``opened the gates'' not only to unify main concepts of theoretical evolutionary biology, theory of computation, and algorithmic information theory, proving the open-ended Darwin-like evolution \cite{Hernandez-Orozco2018,Hernandez-Orozco2018a,Abrahao2015,Chaitin2018}, but also to initiate in \cite{Abrahao2017published,Abrahao2018publishedAMS} a unifying mathematical theoretical framework for complex systems under a both computational and informational perspective, which combines metabiology with
%theoretical biology, theory of computation, algorithmic and 
statistical information theory, game theory, multi-agent systems, and network science. 
In this way, we will briefly discuss in Section~\ref{sectionBackground} how metabiology and algorithmic networks may help bridge abstract formalizations of systemics from mathematical-logical approach in \cite{Bresciani2018} to more realistic models of evolutionary \cite{Hernandez-Orozco2018a} or distributed systems \cite{Prokopenko2009,Fernandez2013,Griffith2014a}.

In the present article, by adding one more important systemic property from the complex systems' ``zoo'' to this theoretical framework, we also introduce an algorithmic-informational measure of synergy and we apply it to our theoretical models, proving that the local algorithmic synergy in solving the halting problem can be as asymptotically large as one may want.

%\section{Motivation}\label{sectionMotivation}

\section{A conceptual background: From systemics, information, and metabiology to algorithmic networks}\label{sectionBackground}

%\subsection{From systemics, information, and metabiology to algorithmic networks}\label{subsectionSystemics}

We start by introducing in this section some of the conceptual background of our results.
For this purpose, we will briefly cross over more abstract approaches to complex systems theory, intertwining foundational subjects and properties related to mathematical logic \cite{Bresciani2018}, information theory \cite{Lizier2018,Griffith2014a} and metabiology \cite{Chaitin2012,Chaitin2018,Hernandez-Orozco2018a,Abrahao2015}, in order to go toward the theory of algorithmic networks \cite{Abrahao2016b,Abrahao2017published,Abrahao2018publishedAMS}.

A system may be initially defined as a unitary entity of a complex and organized nature, made up of a set of active elements \cite{Bresciani2018}. 
It is characterized as a partial structure with functionality.
The general notion of relation used in the definition of system, that of partial relation, is an extension of the usual logical-mathematical concept of relation. 
It was presented by \cite{Mikenberg1986} as basic to the introduction of the mathematical concept of pragmatic truth, later called quasi-truth, and has recently received various applications in logic and philosophy of science (see \cite{Agazzi2010}). 
In this way, it enables one to formally accommodate the incompleteness of information relative to a certain domain of investigation. 

\begin{amsdefinitionundersection}
	 Let $\mathbf{D}$ be a non-empty set. An $n$-ary partial relation $\mathbf{R}$ on $\mathbf{D}$ is an ordered tern 
$ \left< R_1, R_2, R_3 \right> $, 
$R_1$, $R_2$, and $R_3$ with no two of them having common elements, and whose union is $\mathbf{D}$, such that: 
\begin{enumerate}
	\item $R_1$ is the set of the $n$-tuples that we know belong to $\mathbf{R}$; 
	\item $R_2$ is the set of the $n$-tuples that we know do not belong to $\mathbf{R}$; 
	\item $R_3$ is the set of the $n$-tuples for which we do not know if they belong to $\mathbf{R}$ or not.
\end{enumerate} 
\end{amsdefinitionundersection}

\begin{amsdefinitionundersection}
A partial structure is a pair 
$ \left< \mathbf{D}, \mathbf{R}_i \right> $ 
with $\mathbf{D}$ a non-empty set and each 
$ \mathbf{R}_i$, where $ i \in I$, a partial relation on $\mathbf{D}$.
\end{amsdefinitionundersection}

\begin{amsdefinitionundersection}
A system is a partial structure with functionality, that can be denoted by:
\[ \mathbf{S} = \left< \mathbf{D}_i , \mathbf{R}_{ i j } \right>_{ \left( i , j , k \right) \in I \times J \times K }^{ F_k } \] 
$ \mathbf{D}_i $ being the universe of the partial structure, each $ \mathbf{R}_{ i j } $ a partial relation on  $ \mathbf{D}_i $, $F_k$ the functionality, with $I,J,K$ being the respective variation indexes.
\end{amsdefinitionundersection}

Note that a system without the possibility of structural or functional alterations has its universe and functionality constant and can be denoted by $ \mathbf{S} = \left< \mathbf{D} , \mathbf{R}_{ i } \right>_{ i \in I  }^{ F } $.
For instance, this is the case for computable systems in which every transition of internal states with external inputs is a partial recursive function and its functionality is the very computation of this function.

We observe that if $R_3$ does not have any elements ($ R_3 = \emptyset  $),
then $\mathbf{R}$ is a usual $n$-ary relation, so that $ \mathbf{R} = R_1 $, which brings this general definition of a system back to the one in \cite{Abrahao2015}.
Hence, as in \cite{Abrahao2015}, the immediate definition of a computable system, in which $ \mathbf{R} $ is a computable (or recursive) relation and it represents a function, becomes well-defined. 
See also \cite{Hernandez-Orozco2018} for a formalization of this idea in the context of evolutionary systems.
In particular, the reader is invited to note that the oracle-sensitiveness \cite{Abrahao2017published} property of the population directly assures that the respective algorithmic networks always behave exactly like a total function. 
The same also holds for early metabiological models in \cite{Chaitin2012} and for sub-computable or hyper-comnputable versions in \cite{Abrahao2015}.

Functionality is a teleological notion, characterized as a certain informational directioning. 
It may be related to the system’s goals, targets, or ends, and the potential  autonomy of the system’s components may lead to processes which are not individually, but instead globally, self-organized. 
The characteristics of the system may be considered emergences, with systemic synergy, globality and the possibility of novelty being considered among the first properties which appear in the constitution of the very existence of the system. 
In addition, a system is not completely isolated from its environment, because everything, matter, energy, information, that go into or out of the system comes from, passes through, or goes out to the environment. 
For example, in this regard, applications of statistical information theory to stochastic dynamical systems have already shown foundational results in defining and measuring such systemic properties \cite{Lizier2018,Prokopenko2009,Prokopenko2014,Fernandez2013,Oizumi2014}.
However, apart from evolution as in \cite{Chaitin2012,Chaitin2018,Abrahao2015,Hernandez-Orozco2018a}, such study of systemic properties have not been universally applied to deterministic systems.
For this purpose, it leaded us to algorithmic networks, as we will explain below.  

Creation may be the result of transformations conducted by spontaneous and autonomous activities, or from transformations conducted by constitutive and predetermined activities of elements of the system (and eventually boundary elements). 
It may be a new product, or be the result of a process of organizational transformation characterized by the formation of new structures or new functionings. 
In both cases, creation may be thought of as the emergence of a system. 

The process of evolution is characterized as the sequence of states of equilibrium and disequilibrium, manifested in the succession of distinct organizations which arise through the course of transformation of a system. 
If every organization that arises is considered a novelty, then one can affirm that evolution is a sequence of organizational innovations that may be rightly referred to as creative evolution.
In this way, metabiology \cite{Chaitin2012,Chaitin2018,Abrahao2015,Hernandez-Orozco2018a} gave a way to formalize such abstract notions of system, functionality, emergence, and creation in the context of evolutionary systems, whether sub-computable, computable or hyper-computable ones.
In particular, such models show how sole sub-computable, computable, or hypercomputable systems can become more emergently creative (associated to an irreducible increase in the algorithmic information of the system's behavior) under successive random algorithmic mutations and selection of the fittest, which in turn make the environment define the teleological non-intrinsic functionality of the systems as being the increase of the fitness.

Following this same teleology, we showed that one can indeed formally capture the above notions for non-evolutionary systems, giving rise to a formal theory of algorithmic networks \cite{Abrahao2017published}.
For example: 
the model's (non-intrinsic) functionality in \cite{Abrahao2017published} is to increase the average fitnesses of the nodes through the Busy Beaver imitation game (BBIG) at the expense of communication rounds;
and the model's (non-intrinsic) functionality in \cite{Abrahao2018publishedAMS} is to increase the average fitnesses of the nodes through the Busy Beaver imitation game (BBIG) under a Susceptible-Infected-Susceptible contagion scheme at the expense of communication rounds.
Moreover, in this direction, whereas the models in \cite{Abrahao2017published,Abrahao2018publishedAMS} do not formally assign any particular intrinsic common goal to the entire algorithmic network, the present article shows how one can bring the notion of synergistic functionality (translated from statistical-informational measures of synergy in stochastic dynamical systems to networked deterministic systems) to variations of such models.

\section{Preliminary definitions and notation}\label{subsectionPreliminary}

We now restate some main definitions, and notation on which the article results are based.
For a complete introduction to these concepts, see \cite{Wehmuth2016b,Abrahao2017published}.

\subsection{Graphs and networks}\label{subsectionGraphsandnetworks}

\emph{MultiAspect Graphs} (MAGs) $ G $ are generalized representations for different types of graphs \cite{Wehmuth2016b,Wehmuth2017}. 
In particular, a MAG represents dyadic relations between arbitrary $n$-tuples.
Since we aim at a wider range of different network configurations, MAGs allow one to mathematically represent abstract aspects that may appear in complex high order networks \cite{Wehmuth2018b}. For example, these may be dynamic (or time-varying) networks, multicolored nodes or edges, multilayer networks, among others. Moreover, this representation facilitates network analysis by showing that their aspects can be isomorphically mapped into a classical directed graph~\cite{Wehmuth2016b}. Thus, the MAG abstraction has proved to be crucial in~\cite{Abrahao2017published} to establish connections between the characteristics of the network and the properties of the population composed of theoretical machines. Formally, 

\begin{amsdefinitionundersection}\label{defMAG}
	Let $ G=(\mathscr{A},\mathscr{E}) $ be a MultiAspect Graph (MAG), where $\mathscr{E}$ is the set of existing composite edges of the MAG and $\mathscr{A}$ is a class (or list) of sets, each of which is an \emph{aspect}. Each aspect $ \mathbf{ \sigma } \in \mathscr{A} $ is a finite set and the number of aspects $ p = | \mathscr{A} | $ is called the \emph{order} of $ G $. By an immediate convention, we call a MAG with only one aspect as a \emph{first order} MAG, a MAG with two aspects as a \emph{second order} MAG and so on. Each composite edge (or arrow) $ e \in \mathscr{E} $ may be denoted by an ordered $2p$-tuple $ ( a_1,\dots,a_p, b_1, \dots, b_p ) $, where $ a_i, b_i $ are elements of the $i$-th aspect with $ 1 \leq i \leq p = | \mathscr{A} | $. 
\end{amsdefinitionundersection}

$ \mathscr{A}( G ) $ denotes the set (or list) of aspects of $ G $ and $ \mathscr{E}( G ) $ denotes the \emph{composite edge set} of $ G $. We denote the $i$-th aspect of $ G $ as $ \mathscr{A}( G )[i] $. So, $ | \mathscr{A}( G )[i] |$ denotes the number of elements in $ \mathscr{A}( G )[i] $. In order to match the classical graph case, we adopt the convention of calling the elements of the first aspect of a MAG as \emph{vertices}. Therefore, we denote the set $ \mathscr{A}( G )[1] $ of elements of the first aspect of a MAG $ G $ as $ \mathrm{V}(G) $. Thus, a vertex should not be confused with a composite vertex.

Note that, the terms \emph{vertex} and \emph{node} may be employed interchangeably in this article. However, we choose to use the term \emph{node} preferentially in the context of networks, where nodes may realize operations, computations or would have some kind of agency, like in real networks or algorithmic networks. Thus, we choose to use the term \emph{vertex} preferentially in the mathematical context of graph theory.  

Dynamic networks represented by $ G_t =( \mathrm{V}, \mathscr{E}, \mathrm{T} ) $ are time-varying graphs (TVGs) as defined in~\cite{Costa2015a,Wehmuth2016b}. These are a special case of second order MAGs which have only one additional aspect relative to variation over time in respect to the set of nodes/vertices.
Therefore, $\mathrm{V}( G_t )$ is the set of nodes, $\mathrm{T}( G_t )$ is the set of time instants, and $\mathscr{E} \subseteq \mathrm{V}( G_t ) \times \mathrm{T}( G_t ) \times \mathrm{V}( G_t ) \times \mathrm{T}( G_t )$ is the set of edges. Formally:

\begin{amsdefinitionundersection}\label{BdefTVG} 
	Let $ G_t=(\mathrm{V},\mathscr{E},\mathrm{T}) $ be a \emph{time-varying graph} (TVG), where $\mathrm{V}$ is the set of vertices (or nodes), $\mathrm{T}$ is the set of time instants, and $\mathscr{E} \subseteq \mathrm{V} \times \mathrm{T} \times \mathrm{V} \times \mathrm{T}$ is the set of edges\footnote{ That is, the set of existent (second order) composite edges. }.
\end{amsdefinitionundersection}

%We define the set of time instants of the graph $G_t=(\mathrm{V},\mathscr{E},\mathrm{T})$ as $ \mathrm{T}(G_t)=\{t_0, t_1, \dotsc, t_{|\mathrm{T}(G_t)|-1} \} $.
For the sake of simplifying our notations in the theorems below, one can take a natural ordering for $ \mathrm{T}(G_t) $ such that
\[ 
\forall i \in \mathbb{N} \; \left( \, 0 \leq i \leq | \mathrm{T}(G_t) | - 1 \implies t_i = i + 1 \, \right) 
\]

\begin{amsdefinitionundersection}
	Let $d_t(G_t, t_i, u, \tau)$ be the minimum number of time intervals (non-spatial steps or, specially 
	in the present article, node cycles) for a diffusion starting on vertex $u$ at time instant 
	$t_i$ to reach a fraction $\tau$ of vertices in the TVG $G_t$.
\end{amsdefinitionundersection}

In the case the TVG $ G_t $ is connected:

\begin{amsdefinitionundersection}\label{BdefTemporaldiameter}
	Let $D(G_t,t)$ denote the \emph{temporal diffusion diameter} of the TVG $G_t$ taking time instant $t$ as the starting time instant of the diffusion process. That is,
	\[
	D(G_t,t) =
	\begin{cases}
	\infty \quad \text{if} \, \,  \exists u \in \mathrm{V}(G_t) \forall x \in \mathbb{N}  \big( x \neq d_t(G_t,t,u,1) \big) \\
	\max\{ x \mid \, x=d_t(G_t,t,u,1) \, \land \, u \in \mathrm{V}(G_t) \} \quad \text{otherwise}
	\end{cases}
	\]
\end{amsdefinitionundersection}

\begin{amsnotationundersubsection}\label{BdefBinarylogarithm}
	Let $\lg(x)$ denote the binary logarithm $\log_{2}(x)$.
\end{amsnotationundersubsection}

\begin{amsdefinitionundersection}\label{BdefFamilyG_sm}
	Let
	\begin{align*}
	\mathbb{G}_{sm}(f,t,1) = \left\{ G_t \middle| 
	\arraycolsep=1.3pt
	\begin{array}{rl}
	& i = |\mathrm{V}(G_t)| \in \mathbb{N} \\ 
	\land & f(i,t,1) = D(G_t,t)=\mathbf{O}\big( \lg( i ) \big) \, \\ 
	\land & \forall i \in \mathbb{N^*} \exists!G_t \in \mathbb{G}( f,t,\tau )  \left(\;|\mathrm{V}(G_t)|=i \; \right) \\
	\land & \forall u \in \mathrm{V}(G_t) \exists x \in \mathbb{N} \left( \; x = d_t( G_t, t, u, 1) \; \right)  
	\end{array}
	\right\}
	\end{align*}
	\noindent where 
	\[
	\myfunc{f}{ \mathbb{N^*} \times X \subseteq \mathrm{T}(G_t) \times Y \subseteq ]0,1] } { \mathbb{N} } { (x,t,\tau) } { y }
	\]
	\noindent be a \emph{family} of unique sized time-varying graphs that shares $ f(i,t,1) = D(G_t,t) = \mathbf{O}\big( \lg( i ) \big)$, where $i$ is the number of nodes, as a common property.
\end{amsdefinitionundersection}

\subsection{Formal languages, machines, and algorithmic information theory}\label{subsectionAIT}

\begin{Bnotation}
	Let $ (x)_2 $ denote the binary representation of the number $ x \in \mathbb{N} $. 
	In addition, let $ (x)_{L} $ denote the representation of the number $ x \in \mathbb{N} $ in language $ L $.
	Analogously, let $ (w)_{10} \in \mathbb{N}$ denote the decimal representation of the string $ w \in L $, where $ w = \left( (w)_{ 10 } \right)_L $.
\end{Bnotation}

\begin{Bnotation}\label{BdefFunctionphi}
	Wherever number $ n \in \mathbb{N} $ appears in the domain or in the codomain of a partial (or total) function
	\[
	\myfunc{ \varphi_{ \mathcal{U} } }{ L }{ L }{ x }{ y = \varphi_{ \mathcal{U} }(x) } \text{ ,}
	\]
	\noindent where $ \mathcal{U} $ is a Turing machine, or an oracle Turing machine, running on language $L$, it actually denotes
	\[
	\left( n \right)_{ L }
	\]
\end{Bnotation}

\begin{amsdefinitionundersection}\label{BdefHaltingtimefunction}
	Let 
	$ \myfunc{ T }{ \, \mathbf{L_U} \times \mathbf{L_U} }{ \mathbb{N} }{ \left( \mathrm{M} , p \right) }{ T( \mathrm{M} , p ) = n } $ 
	be the \emph{partial} recursive function that returns the computation time that machine $ \mathrm{M} $ takes to halt on input $ p $.
\end{amsdefinitionundersection}

As in \cite{Abrahao2016,Chaitin2012},
\begin{amsdefinitionundersection}\label{BdefFunctionBB}
	Let 
	\[ \myfunc{ BB }{ \left\{ n \; \middle\vert \; \min\left\{ \left| s \right| \, \vert \, s \in \mathbf{L_U} \land \exists x \left(  \mathbf{U}\left( s \right) = x \right) \right\}  \leq n \right\} \subseteq \mathbb{N} }{ \mathbb{N} }{ n }{ BB(n) = k } \]
	be the total\footnote{ Without loss of generality, one can choose a universal self-delimiting programming language $ \mathbf{L_U} $ in which there is $ x \in \mathbf{L_U} $ such that $ \mathbf{U}\left( w \right) = x $ and $ \left| w \right| = \min\left\{ \left| s \right| \; \vert \, s \in \mathbf{L_U} \right\} $, where $ w \in \mathbf{L_U} $.} function that returns the largest integer that a program $ p \in \mathbf{L_U}  $ with length $ \leq n \in \mathbb{N} $ can output running on machine $ \mathbf{U} $. More formally:
	\[
	BB(n) = \max\left\{ i \; \middle\vert \; \mathbf{U}\left( p \right) = w \, \land i = \left( w \right)_{10} \, \land \, \left| p \right| \leq n \right\}
	\] 
\end{amsdefinitionundersection}

\begin{amsnotationundersubsection}\label{notationPairing}
	Let $ \left< \, \cdot \, , \, \cdot \, \right> $ denote an arbitrary recursive bijective pairing function. This notation can be recursively extended to $ \left<   \, \cdot \, ,  \left< \, \cdot \, , \, \cdot \, \right> \right> $ and, then, to an ordered tuple $ \left< \, \cdot \, , \, \cdot \,  \, , \, \cdot \,\right> $. This iteration can be recursively applied with the purpose of defining finite ordered tuples $ \left< \cdot \, , \, \dots \, , \, \cdot   \right> $.
\end{amsnotationundersubsection}

\begin{Bnotation}\label{BdefConcatenation}
	Let $ \mathrm{\textbf{L}}_{\mathbf{U}} $ denote a recursive binary self-delimiting (or prefix-free) universal programming language for a universal Turing machine $\mathbf{U}$ such that there is a concatenation of strings $w_1, \dots , w_k$ in the language $ \mathrm{\textbf{L}}_{\mathbf{U}} $, which preserves\footnote{ For example, by adding a prefix to the entire concatenated string $ w_1 w_2  \dots  w_k $ that encodes the number of concatenations. Note that each string was already self-delimiting. See also \cite{Abrahao2016,Abrahao2017published} for more discussions and properties of the notation ``$ \circ $''. } the self-delimiting (or prefix-free) property of the resulting string, denoted by 
	\[
	w_1 \circ  \dots  \circ w_k \in \mathrm{\textbf{L}}_{\mathbf{U}}
	\]
	In addition, $  \mathbf{L_U} $ is a complete binary code with
	\[
	\sum\limits_{ p \in \mathbf{L_U} } \frac{1}{ 2^{ | p | } } = 1
	\]
	
\end{Bnotation}

The reader may also note that this self-delimiting-preserving concatenation ``$ \circ $'' is just one example of recursive pairing bijective function $ \left< \cdot \, , \, \cdot \right> $. 
%We know that this pairing function can be extended to $ \left<   \, \cdot \, , \, \left< \, \cdot \, , \, \cdot \, \right> \right> $ and, then, to an ordered triple $ \left< \, \cdot \, , \, \cdot \,  \, , \, \cdot \,\right> $. 
%This way, this procedure can be recursively applied with the purpose of defining finite ordered tuples $ \left< \cdot \, , \, \dots \, , \, \cdot   \right> $. 
In addition, choosing between two distinct recursive pairing bijective functions $ \left< \cdot \, , \, \cdot \right>_1 $ and $ \left< \cdot \, , \, \cdot \right>_2 $, can only affect the algorithmic complexity\footnote{ See Definition~\ref{BdefAlgComp}. } by
\[
A( \left< w_1 \, , \, w_2 \right>_1 ) = A( \left< w_1 \, , \, w_2 \right>_2 ) \pm \mathbf{O}(1)
\] 
\noindent Therefore, the reader may equivalently replace\footnote{ Along with the appropriate re-interpretation of what is prefixes or suffixes in language $ \mathbf{L_U} $. }
\[
w_1 \circ \dots \circ w_k
\]
\noindent with
\[
\left< w_1 \, , \, \dots \, , \, w_k \right>
\]
\noindent in the present article without affecting the final result.

\begin{amsdefinitionundersection}\label{BdefProgramp_t}
	Let $ \mathrm{p}_{T} \in \mathbf{L_U} $ be any program of $ \mathbf{U} $ that computes a partial recursive function such that
	\[
	\mathbf{U}\left( \mathrm{p}_T \circ p \right) = \left(T( \mathbf{U} , p )\right)_{\mathbf{L_U}}
	\]
	\noindent where function $ T( \mathrm{M} , p ) $ holds as in Definition~\ref{BdefHaltingtimefunction}. 
\end{amsdefinitionundersection}

\begin{amsdefinitionundersection}\label{BdefFunction+1}
	Let $ \mathrm{p}_{+1} \in \mathbf{L_U} $ be any program of $ \mathbf{U} $ that computes a partial recursive function such that
	\[
	\mathbf{U}\left( \mathrm{p}_{+1} \circ p \right) = { \left( { \left(  \mathbf{U}\left( p \right) \right) }_{10} + 1 \right) }_{ \mathbf{L_U} }
	\]
\end{amsdefinitionundersection}

\begin{amsdefinitionundersection}\label{BdefAlgComp}
	The (unconditional) \emph{prefix} \emph{algorithmic complexity} (also known as self-delimiting program-size complexity or Solomonoff-Komogorov-Chaitin complexity) of a finite binary string $ w $, denoted by $ A(w) $, is the length of the shortest program $w^* \in \mathbf{L_U} $ such that $ \mathbf{U}(w^*) = w $.\footnote{ $ w^* $ denotes the lexicographically first $ \mathrm{p} \in \mathbf{L_U} $ such that $ \left| \mathrm{p} \right| $ is minimum and $ \mathbf{U}(p) = w $.} The \emph{conditional} prefix algorithmic complexity of a binary finite string $ y $ given a binary finite string $ x $, denoted by $ A( y \, | x ) $, is the length of the shortest program $w^* \in \mathbf{L_U} $ such that $ \mathbf{U}( \left< x , w^* \right> ) = y $. Note that $ A( y ) = A( y \, | \epsilon ) $, where $ \epsilon $ is the empty string. Similarly, we have the \emph{joint} prefix algorithmic complexity of strings $x$ and $y$ defined by $ A( x , y ) = A( \left< x , y \right> ) $ or $ A( x , y ) = A(  x \circ y  ) $, the \emph{prefix algorithmic complexity of information} in $x$ about $y$ denoted by $ I_K( x : y ) = A(y) - A( y \, | x ) $, and the \emph{mutual algorithmic information} of the two strings $x$ and $y$ denoted by $ I_A( x \, ; y ) = A(y) - A( y \, | x^* ) $.
\end{amsdefinitionundersection}

%The reader may also find in the literature the prefix algorithmic complexity in Definition~\ref{BdefAlgComp} also denoted by $H(w)$ or $K(w)$ \cite{Calude2002,Chaitin2004,Li1997,Downey2010}. 
%In fact, from $ A(w) $, one can define $ I_A( w ) $ as the algorithmic information contained in a object about itself. To this end, note that we have that, if $ A(w) $ denotes the prefix algorithmic complexity of $w$, $ A( w' | w ) $ denotes the conditional prefix algorithmic complexity of $w'$ given $w$, $ I_K( w : w' ) = A(w') - A( w' | w ) $ denotes the K-complexity of information in $w$ about $w'$, and $ I_A( w ; w' ) = A(w') - A( w' | w^* )  $ denotes the mutual algorithmic information of two objects $w$ and $w'$, then $ I_K( w : w ) = A( w ) - \mathbf{O}(1) $ and $ I_A( w ; w ) = A(w) - \mathbf{O}(1)  $. Thus, the reader may also choose to define $ I_A( w ) $ as an equivalence class in which
%	\begin{gather*}
%	\left| I_A( w ) -  I_K( w : w ) \right|=\mathbf{ O }(1) \\
%	\text{and} \\
%	\left| I_A( w ) -  I_A( w ; w ) \right|=\mathbf{ O }(1) \\
%	\end{gather*}
%\noindent This way, we will have that $ \left| I_A( w ) -  A(w) \right|=\mathbf{ O }(1)  $ and, therefore, the results of the present article hold anyway for employing interchangeably both $ I_A( w ) $ or $ A(w) $.

Now, since we will be dealing with a population of randomnly generated arbitrary programs (i.e., Turing machines) in Section~\ref{sectionModel}, we need to define a theoretical machine that can return fitness values for both halting and non-halting programs:

\begin{amsdefinitionundersection}\label{BdefU'}
	Let $ \mathbf{L_U} $ be the recursive binary self-delimiting universal programming language $ \mathrm{\textbf{L}}_{\mathbf{U}} $ (as in Notation~\ref{BdefConcatenation}) for a universal Turing machine $\mathbf{U}$, where there is a constant $ \epsilon \in \mathbb{R} $, with $ 0 < \epsilon \leq 1 $, and a constant $  C_{L} \in \mathbb{N} $ such that, for every $ N \in \mathbb{N} $, 
	\[
	A(N) \leq \lg(N) + ( 1+\epsilon )\lg(\lg(N)) + C_{L}
	\] 
	\noindent We define an \emph{oracle\footnote{ Or any hypercomputer with a respective Turing degree higher than or equal to $\mathbf{0'} $.} Turing machine} $\mathbf{U'}$ such that, for every $w \in \mathrm{\textbf{L}}_{\mathbf{U}}$,
	\[ \mathbf{U'}(w) =
	\begin{cases}
	{\mathbf{U}}(w) \text{``}+ 1\text{''} & \quad \text{if  } \mathbf{U} \text{ halts on } w \\
	\text{``} 0 \text{''} & \quad \text{if  } \mathbf{U} \text{ does not halt on } w \\
	\end{cases}
	\]
	
\end{amsdefinitionundersection}

Note that, since $ \mathbf{L_U} $ is self-delimiting and $ {\mathbf{U}}(w) \in \mathbf{L_U} $, we have that the operator $ \text{``}+1\text{''} $ actually means the successor operator in an arbitrary recursive enumeration of language $ \mathbf{L_U} $. In the same manner, we have that $ \text{``} 0 \text{''} $ actually means $ \left( 0 \right)_{ \mathbf{L_U} } $.

Thus, the oracle Turing machine in Definition~\ref{BdefU'} is basically (except for a trivial 
bijection) the same as the chosen universal Turing machine. The oracle is only triggered to 
know whether the program halts or not in first place.

\section{Previous work on algorithmic networks}

In this section, we remember the previous work \cite{Abrahao2017published,Abrahao2018publishedAMS,Abrahao2018c,Abrahao2016b} on which this article is based.

\subsection{Algorithmic networks}\label{subsectionAN}

We remember here the general definition of \emph{algorithmic networks} $ \mathfrak{N} = (G, \mathfrak{P}, b)$ in~\cite{Abrahao2017published}. It is a triple $ (G, \mathfrak{P}, b) $ defined upon a population of theoretical machines $\mathfrak{P}$, a generalized graph $G=(\mathscr{A},\mathscr{E})$, and a function $b$ that makes aspects of $G$ to correspond to properties of $\mathfrak{ P }$, so that a node in $\mathrm{V}(G)$ is mapped one-to-one to an element of $\mathfrak{ P }$. Formally:

\begin{amsdefinitionundersection}\label{BdefAN}
	We define an \emph{algorithmic network} $ \mathfrak{N} = (G, \mathfrak{P}, b)$ upon a population of theoretical machines $\mathfrak{P}$, a MultiAspect Graph $G=(\mathscr{A},\mathscr{E})$ and a function $b$ that causes aspects of $G$ to be mapped\footnote{ See Definition \ref{BdefFunctionbinAN} .} into properties of $\mathfrak{ P }$, so that a vertex in $\mathrm{V}(G) $ corresponds one-to-one to a theoretical machine in $\mathfrak{ P }$ and the communication channels through which nodes can send or receive information from its neighbors are defined precisely by composite edges (or, for directed graphs, composite arrows) in
	$ G $. 
\end{amsdefinitionundersection}

We define a \emph{population} $\mathfrak{P}$ as an ordered sequence\footnote{ In which repetitions are allowed.} $  \left( o_1 , \dots , o_i , \dots , o_{ \left| \mathfrak{P} \right| } \right) $, where $ X $ is the support set of the population and $f_o$ is a labeling surjective function
\[
\myfunc{ f_o }{ \mathfrak{ P } = \left( o_1 , \dots , o_i , \dots , o_{ \left| \mathfrak{P} \right| } \right)  }{ X \subseteq L }{ o_i }{ f_o( o_i ) = w } \text{,}
\]
\noindent where $L$ is the language on which the chosen theoretical machine $U$ are running. Each member of this population may receive inputs and return outputs through communication channels. A \emph{communication channel} between a pair of elements from $\mathfrak{P}$ is defined in $\mathscr{E}$ by a composite edge (whether directed or undirected) linking this pair of nodes/programs.

Third, we define function $b$ as

\begin{amsdefinitionunderamsdefinitionundersection}\label{BdefFunctionbinAN}
	Let 
	\[ \myfunc{b}{ Y \subseteq \mathscr{A}(G) } { X \subseteq Pr(\mathfrak{P}) } { \mathbf{\overline{a}}  } { b( \mathbf{\overline{a}} ) = \mathbf{\overline{p_r}} } \]
	be a function that \emph{maps} a subspace of aspects $Y$ in $\mathscr{A}$ into a 
	subspace of properties $X$ in the set of properties $Pr(\mathfrak{P})$ of the respective 
	population such that there is a bijective function $ f_{V\mathfrak{ P }} $ such that, for 
	every $ (v,\mathbf{ 
		\overline{x} }) \in Y \subseteq \mathscr{A}(G) $ with $ b( v,\mathbf{ \overline{x} } ) = ( o_i, 
	b_{ dim( Y ) - 1 }( \mathbf{ \overline{x} } ) ) \in X$,
	\[ \myfunc{f_{V\mathfrak{ P }}}{ \mathrm{V}(G)  } { \mathfrak{P} = \{ o_i \mid f_o( o_i ) = 
		w \in L \} }   { v } { f_{V\mathfrak{ P }}(v) = o_i } \text{ ,} \]
	\noindent where $v$ is a vertex (or node) and $o_i$ is an element of the 
	sequence/population $\mathfrak{ P }$.
\end{amsdefinitionunderamsdefinitionundersection}

We say an element $ o_i \in \mathfrak{P} $ is \emph{networked} \textit{iff} there is $ 
\mathfrak{N} = (G, \mathfrak{P}, b) $ such that $o_i$ is running as a node of $ \mathfrak{N} 
= (G, \mathfrak{P}, b) $, where $ \mathscr{E}(G) $ is non-empty\footnote{ That is, there must be at least one composite edge connecting two elements of the algorithmic network. }.
We say $o_i$ is \emph{isolated} otherwise. That is, it is only functioning as an element of $ \mathfrak{P} $ and not as a node of $ \mathfrak{N} = (G, \mathfrak{P}, b) $.
We say that an input $ w \in L $ is a \emph{network input} \textit{iff} it is the only external source of information every node/program receives and it is given to every node/program before the algorithmic network begins any computation. 
Note that letter $w$ may also appear across the text as denoting an arbitrary element of a language. It will be specified in the assumptions before it appears or in statement of the respective definition, lemma, theorem or corollary.

\begin{amsdefinitionundersection}\label{BdefCycle} 
	A \emph{node cycle} in a population $ \mathfrak{ P } $ is defined as a node/program returning an \emph{output} (which, depending on the language and the theoretical machine the nodes are running on, is equivalent to a node completing a halting computation).
	\begin{enumerate}
		\item If this node cycle is \emph{not} the last node cycle, then its respective output is 
		called a \emph{partial output}, and this partial output is shared (or not, which depends on 
		whether the population is networked or isolated) with the node's neighbors, accordingly 
		to a specific information-sharing protocol (if any);
		\item If this node cycle is the last one, then this output is called a \emph{final output} such that no more information is shared through the network; 
		\item If every node/program in $ \mathfrak{ P } $ has completed its last node cycle, returning its final outputs, and the population $ \mathfrak{ P } $ is running networked by an algorithmic network $ \mathfrak{N} = (G, \mathfrak{P}, b) $, then we say the algorithmic network $ \mathfrak{N} $ as a whole completed an \emph{algorithmic network cycle}.
	\end{enumerate}
	In addition, let 
	\[
	\mathfrak{C} = \bigcup_{ o_i \in \mathfrak{ P } }  \mathfrak{C}( o_i )
	\]
	be the set of the maximum number of node cycles that any node/program $o_i$ in the population $ \mathfrak{P} $ would be able to perform in order to return a final output, where $\mathfrak{C}(o_i) \subseteq \mathbb{N} $ is the set of all node cycles that node/program $o_i$ can perform.
\end{amsdefinitionundersection}

In the particular case a population is defined on the language $ \mathbf{L_U} $ and machine $ 
\mathbf{U'} $ we assume the 
notation:

\begin{amsdefinitionundersection}\label{BnotationPartialoutputs}
	Let $ p_{net_{ \mathbf{U} } } $ be a program such that $ p_{net_{ \mathbf{U} } } \circ o_i \circ c $ computes on machine $ \mathbf{U'} $ cycle-by-cycle what a node/program $o_i \in \mathfrak{P} $ does on machine $ \mathbf{U} $ until cycle $c$ when networked by $ \mathfrak{N} = (G, \mathfrak{P}, b)$. 
	Let $ p_{iso_{ \mathbf{U} } } $ be a program such that $ p_{iso_{ \mathbf{U} } } \circ o_i \circ c $ computes on machine $ \mathbf{U'} $ cycle-by-cycle what a node/program $o_i \in \mathfrak{P} $ does on machine $ \mathbf{U} $ until cycle $c$ when isolated. 
	Let $p_{o_i,c}$ be the \emph{partial output} sent by node/program $o_i$ at the end of cycle $c$. Also, $p_{o_i, \max\{ c \mid c \in \mathfrak{C}( o_i ) \}}$ denotes the final output of the node/program $o_i$.
\end{amsdefinitionundersection}

\begin{amsdefinitionundersection}\label{BnotationNeighborset}
	Let $\mathbf{X}_{neighbors}(o_j,c)$ be the set of incoming neighbors of node $o_j$ that have sent partial outputs to it at the end of the cycle $c$ when running on algorithmic networked $ \mathfrak{N} = (G, \mathfrak{P}, b)$.
	Let $ \{ p_{o_i,c} \mid o_i \in \mathbf{X}_{neighbors}(o_j, c) \land i \in \mathbb{N} \land c \in 
	\mathfrak{C} \} $ be the set of partial outputs relative to $\mathbf{X}_{neighbors}(o_j,c)$.
\end{amsdefinitionundersection}

\subsection{ Busy Beaver imitation game }\label{subsectionBBIG}

In \cite{Abrahao2017published}, we have narrowed our theoretical approach by defining a class of algorithmic networks $  {\mathfrak{N}_{BB}}_{sm}(N,f,t,1,j) $---also denoted by a triplet as $ ( G_t, \mathfrak{P}_{BB} (N), b_j ) $---in which their populations $   {\mathfrak{P}_{BB}}(N)  $ and TVGs $ G_t \in \mathbb{G}_{sm}(f,t,1) $ have determined properties. 

As defined in Section~\ref{subsectionAN}, each element of the population corresponds one-to-one to a node/vertex in $ G_t $ and each time instant in $ G_t $ is mapped to a cycle (or communication round). These mappings are defined by the function $ b_j $. 

The population $  {\mathfrak{P}_{BB}}(N) $ is composed of randomly generated Turing machines (or randomly generated self-delimiting programs) which are represented in a self-delimiting universal programming language $ \mathbf{L_U} $. This population is synchronous with respect to halting cycles, that is, in the end of a cycle (or communication round, as in distributed computing) every node returns its outputs at the same time. Nodes that do not halt in any cycle always return as final output the lowest fitness, that is, the integer value $ 0 $. Here, a straightforward interpretation is that nodes that eventually do not halt in a cycle are ``killed'', so that their final output has the ``worst'' fitness.  Thus, these nodes are programs that ultimately run on an oracle Turing machine (or a hypercomputable system) $ \mathbf{U'} $---this requirement is also analogous to the one in \cite{Abrahao2016,Chaitin2014,Hernandez-Orozco2018}, which deal with a sole program at the time and not with a population of them. However, the oracle is only necessary to deal with the non-halting computations.  That is, $ \mathbf{U'} $ behaves like an universal Turing machine $ \mathbf{U} $ except that it returns zero whenever a non-halting computation occur. 

In addition, the networked population $  {\mathfrak{P}_{BB}}(N)  $ follows an \emph{imitation-of-the-fittest protocol} (IFP), diffusing the information of the fittest randomly generated node (i.e., the node that partially outputs the largest integer in cycle $1$)\footnote{ As in~\cite{Abrahao2016b,Chaitin2014,Chaitin2018,Hernandez-Orozco2018}, note that we still use the Busy Beaver function as a complexity measure for fitness. Therefore, the largest integer directly represents the fittest final output of a node.}. Thus, every node in $ {\mathfrak{N}_{BB}}_{sm}(N,f,t,1,j) $ obeys the IFP, in which after the first cycle (i.e., after the first round of partial outputs) every node only imitates the neighbor that has partially output the largest integer, repeating this value as its own partial output in the next cycle. Thus, the main idea defining the IFP is a procedure in which each node $o_i$ compares its neighbors' partial output (that is, the integer they have calculated in the respective cycle) and runs the program of the neighbor that have output the largest integer if, and only if, this integer is larger than the one that the very node $o_i$ has output. Since $ {\mathfrak{N}_{BB}}_{sm}(N,f,t,1,j) $ is playing the Busy Beaver game \cite{Abrahao2017published} on a network while limited to simple imitation performed by a randomly generated population of programs, we say it is playing a \emph{Busy Beaver Imitation Game} (BBIG). A (network) Busy Beaver game \cite{Abrahao2017published} is a game in which each player is trying to calculate the largest integer---as established as our measure of fitness or payoff\footnote{ See \cite{Abrahao2017published,Abrahao2016,Hernandez-Orozco2018} for more discussions.}---it can using the information shared by its neighbors. Thus, the BBIG is a special case of the Busy Beaver game.

Thus, $ {\mathfrak{N}_{BB}}_{sm}(N,f,t,1,j) $ is a synchronous algorithmic network populated by $N$ randomly generated nodes such that, after the first cycle (or arbitrary $c_0$ cycles), it starts a diffusion process of the biggest partial output (given at the end of the first cycle) determined by network $ G_t $: at the first time instant each node may receive a network input $w$, which is given to every node in the network, and runs separately (i.e., not networked), returning its respective first partial output; then, the plain diffusion of large integers starts as determined by the IFP through the respective dynamical network $G_t$. At the last time instant contagion stops and one cycle (or more) is spent in order to make each node to return a final output. Formally,

\begin{amsdefinitionundersection}\label{BdefN_BB}
	Let
	\[
	{\mathfrak{N}_{BB}}_{sm}(N,f,t,1,j) = (G_t, \mathfrak{P}_{BB} (N),b_j)
	\]
	\noindent be an algorithmic network, where $f$ is an arbitrary well-defined function such that 
	\[
	\myfunc{f}{ \mathbb{N^*} \times X \subseteq \mathrm{T}(G_t) \times Y \subseteq ]0,1] } { \mathbb{N} } { (x,t,\tau) } { y }
	\]
	\noindent and $ G_t \in \mathbb{G}_{sm}(f,t,1) $, $ | \mathrm{V}(G_t) | = N $, $ | \mathrm{T}(G_t) | > 0 $, and there are arbitrarily chosen\footnote{ Since they are arbitrarily chosen, one may choose to take them as minimum as possible in order to minimize the number of cycles for example. That is, $ c_0 = 0 $ and $ n = | \mathrm{T}(G_t) | + 1 $ for example. } $ c_0, n \in \mathbb{N} $ with $ c_0 + | \mathrm{T}(G_t) | + 1 \leq n \in \mathbb{N} $ such that $b_j$ is an injective function, where
	\[
	\myfunc{b_j} { \mathrm{V}(G_t) \times \mathrm{T}(G_t) } { \mathfrak{P}_{BB}(N) \times \mathbb{N}|_1^{ n }} { (v,t_{c-1}) } { b_j(v,t_{c-1})=( o_i , c_0 + c ) }  
	\]
	
\end{amsdefinitionundersection}

Since the way time instants are mapped into cycles is fixed given values of $c_0$ and $n$, we may equivalently denote function $ b_j $ as
\[
\myfunc{b_j} { \mathrm{V}(G_t) } { \mathfrak{P}_{BB}(N) } { v } { b_j( v )=( o_i ) }  
\]

\subsection{Background results}\label{subsectionBackgroundresults}

Following an algorithmic approach to evolutionary open-endedness (EvoOE), we have found in \cite{Abrahao2017published} that open-endedness may also emerge as an akin, but different, phenomenon to EvoOE: Instead of achieving an unbounded quantity of algorithmic complexity over time (e.g., after successive mutations), an unbounded quantity of emergent algorithmic complexity is achieved as the population/network size increases indefinitely. The algorithmic complexity of a node/program's final output when networked minus the algorithmic complexity of a node/program's final output when isolated formally defines an irreducible quantity of information that \emph{emerges} in respect to a node/program that belongs to an algorithmic network. We call it as \emph{emergent algorithmic complexity (EAC)} of a node/program. The reader may also find more discussions on emergence and open-endedness in \cite{Abrahao2017published,Abrahao2018publishedAMS}.

Formally, we have defined \emph{average emergent open-endedness} in the context of general algorithmic networks as

\begin{amsdefinitionundersection}\label{BdefAEOE}
	We say an algorithmic network $ \mathfrak{N} $ with a population of $N$ nodes has the property of \emph{average (local) emergent open-endedness (AEOE)} for a given network input $w$ in $c$ cycles \textit{iff} 
	\[
	\lim_{ N \to \infty } \mathbf{E}_{ \mathfrak{N} } 
	\left(
	{ {\displaystyle{\myDelta_{iso}^{net}} A} (o_i,c)} 
	\right)
	=
	\infty
	\]  
\end{amsdefinitionundersection}

\noindent And, in the case of an algorithmic network $ \mathfrak{N} = (G, \mathfrak{ P }, b) $ with randomly generated nodes, we call this property as \emph{expected (local) emergent open-endedness}. We have that

\begin{align*}
&\mathbf{E}_{ \mathfrak{N} } 
\left(
{ {\displaystyle{\myDelta_{iso}^{net}} A} (o_i,c)} 
\right)
= \\
&=
{\tiny 
	\sum\limits_{ b }
}
\frac{
	\frac{
		\sum\limits_{ o_i \in \mathfrak{P} } { {\displaystyle{\myDelta_{iso}^{net(b)}} A} (o_i,c)}
	}
	{N} 
}
{ | \{ b \} | } 
\end{align*}

\noindent denotes the \emph{average emergent algorithmic complexity of a node/program (AEAC)} in an algorithmic network $ \mathfrak{N} = (G, \mathfrak{ P }, b) $ with network input $w$. In addition:

\begin{amsdefinitionundersection}\label{BdefEAC}
	The \emph{emergent algorithmic complexity (EAC)} of a node/program $o_i$ in $c$ cycles is given in an algorithmic network that always produces partial and final outputs by

	\[
	{\displaystyle{\myDelta_{iso}^{net(b)}}} A (o_i,c) = 
	A (\mathbf{U}(p_{net}^{b} ( o_i , c )) - 
	A (\mathbf{U}(p_{iso} ( o_i , c) )
	\]
	
	\noindent where:
	\begin{enumerate}
		\item $ f_o( o_i ) \in L $;
		
		\item $ p_{net}^{b} ( o_i , c ) $ represents the program that returns the final output of $o_i$ when networked assuming the position $ v $, where $ b(v,\mathbf{\bar{x}}) = (o_i, b_{ dim( Y ) - 1 }( \mathbf{ \overline{x} } ) ) $, in the MAG $G$ in the specified number of node cycles $c$ with network input $w$; 
		
		\item $ p_{iso} ( o_i , c)  $ represents the program that returns the final output of $o_i$ when isolated in the specified number of node cycles $ c $ with network input $w$;
	\end{enumerate}
	
\end{amsdefinitionundersection}

Note that the network input $ w $ was omitted in $ p_{net}^{b} ( o_i , c ) $ and  $ p_{iso} ( o_i , c)  $, as presented in \cite{Abrahao2017published,Abrahao2018publishedAMS}.
This is because in the models in \cite{Abrahao2017published,Abrahao2018publishedAMS} we were focusing on lower bounds on the expected emergent algorithmic complexity of a node and it was achieved by estimating the occurrence of fittest node/programs that ignores its inputs.
However, for the following results in this article, we may equivalently denote $ p_{net}^{b} ( o_i , c ) $ as $ p_{net}^{b} ( o_i , c , w ) $ and $ p_{iso} ( o_i , c)  $ as $ p_{iso} ( o_i , c , w )  $.

In \cite{Abrahao2017published}, we showed that there is a lower bound for the expected emergent algorithmic complexity in algorithmic networks ${\mathfrak{N}_{BB}}_{sm}$ such that it depends on how much larger is the average diffusion density (in a given time interval) $ { \tau_{\mathbf{E}(max)}( N,f,t, 1 ) }|_{t}^{t'} $ compared to the cycle-bounded conditional halting probability $ \Omega(w,c(x))  $. Formally:

\begin{thmundersection} \label{BthmMainsmalldiameter} 
	Let $ w \in \mathbf{L_U} $ be a network input. 	
	Let $ 0 < N \in \mathbb{N} $.
	Let ${\mathfrak{N}_{BB}}_{sm}(N,f,t,1,j) = ( G_t, \mathfrak{ P }_{BB}(N), b_j ) $ be well-defined.
	Let $ t_0 \leq t \leq t' \leq t_{ |\mathrm{T}(G_t)|-1 } $.
	Let 
	\[ \myfunc{c}{ \mathbb{N} } { \mathfrak{C_{BB}} } { x } { c(x)=y } \]
	\noindent be a total computable function where $ c(x) \geq c_0 + t'+1 $. 
	Then, we will have that:
	%s25p13l1
	\begin{align*}
	& \lim\limits_{ N \to \infty }
	\mathbf{E}_{{\mathfrak{N}_{BB}}_{sm}(N,f,t,1) } 
	\left(
	{ {\displaystyle{\myDelta_{iso}^{net}} A} (o_i,c(x))} 
	\right)
	\geq
	\lim\limits_{ N \to \infty }
	\left( 
	{ \tau_{\mathbf{E}(max)}( N,f,t,1 ) }|_{t}^{t'}
	-
	\Omega(w,c(x))
	\right)
	\lg(N) - \\
	& - \Omega(w,c(x)) \lg(x) - 2 \, \Omega(w,c(x))\lg(\lg(x)) - A(w) - C_5
	\end{align*}
\end{thmundersection}

This lower bound also depends on the parameter $ x $ for which one is calculating the number of node cycles. In fact, we have proved that our results hold even in the case of spending a computably larger number of node cycles compared to $ x $. Furthermore, we have proved that the small-diameter phenomenon is a condition that ensure that there is a central time $ t_{cen_1}(c) $ to trigger expected emergent open-endedness. Formally:

\begin{thmundersection} \label{BthmMainCentralTimesmalldiameter}
	Let $ w \in \mathbf{L_U} $ be a network input.
	Let $ 0 < N \in \mathbb{N} $.
	If there exist $ 0 \leq z_0 \leq | \mathrm{T}(G_t) | -1 $ and $ \epsilon, \, \epsilon_2 > 0 $ such that 
	\[
	z_0 + f( N, t_{z_0}, 1 ) + 2 
	= 
	\mathbf{ O }
	\left( \frac
	{ N^{ C } }
	{ \lg(N) } 
	\right)
	\]
	\noindent where
	\[ 
	0
	<
	C = 
	\frac
	{
		{ \tau_{\mathbf{E}(max)}( N,f,t_{z_0},1 ) }|_{ t_{z_0} }^{ t_{ z_0 + f( N, t_{z_0}, 1 ) }  }
		-
		\Omega(w, c_0 + z_0 + f( N, t_{z_0}, 1 ) + 2 )
		-
		\epsilon
	}
	{ \Omega(w,  c_0 + z_0 + f( N, t_{z_0}, 1 ) + 2  ) }
	\leq
	\frac{1}{ \epsilon_2 }
	\]
	Then, for every non-decreasing total computable function $ \myfunc{c}{ \mathbb{N} } { \mathfrak{C_{BB}} } { x } { c(x)=y } $, where $ t_{z_0}, \, t_{ z_0 + f( N, t_{z_0}, 1 ) } \in \mathrm{T}(G_t) $ and $ c(z_0 + f( N, t_{z_0}, 1 ) + 2) \geq c_0 + z_0 + f( N, t_{z_0}, 1 ) + 2 $ and $ {\mathfrak{N}_{BB}}_{sm} (N,f,t_{z_0},1,j) = ( G_t, \mathfrak{ P }_{BB}(N), b_j ) $ is well-defined, we will have that there is $ t_{cen_1}(c) $ such that $ t_{cen_1}(c) \leq t_{ z_0 } $.
\end{thmundersection}

Our proofs follow mainly from information theory, computability theory, and graph theory.  Therefore, we have shown that there are topological conditions (e.g., the small-diameter phenomenon) that trigger a phase transition in which eventually the algorithmic network $ {\mathfrak{N}_{BB}}_{sm}  $ begins to produce an unlimited amount of bits of average local emergent algorithmic complexity/information. These conditions come from a positive trade-off between the average diffusion density and the number of cycles (i.e., communication rounds). Thus, the diffusion power of a dynamic (or static) network has proved to be paramount with the purpose of optimizing the average fitness/payoff of an algorithmic network that plays the Busy Beaver imitation game in a randomly generated population of Turing machines.

\section{A model of algorithmic network for synergistically solving a common problem}\label{sectionModel}

In this section, we present the model of algorithmic networks on which we will prove lemmas and theorems. 
In this way, we focus on the description and the definition of the new model. 

The main idea that defines the algorithmic networks $ \mathfrak{N}_{SBB} $ is to formalize a distinct information-sharing (or communication) protocol that is based on the BBIG, so that nodes can use the largest integer the other nodes are calculating to nourish a global procedure in order to compute a function. 
So, this is a modification of the algorithmic networks $ {\mathfrak{N}_{BB}}_{sm} $ in \cite{Abrahao2017published}.
The latter only follow the IFP with a plain spreading of the largest integer, as described in Section~\ref{subsectionBBIG}.
In particular, with respect to $ {\mathfrak{N}_{BB}}_{sm} $ in \cite{Abrahao2017published}, we will modify the functioning of the IFP in the last node cycle, just in the moment each node is about to return its final output.
First, we will describe the properties of  $ \mathfrak{N}_{SBB} $ that are common to $ {\mathfrak{N}_{BB}}_{sm} $.
Then, we will describe the functioning of the \emph{synergistic imitation-of-the-fittest protocol} (SIFP) that is different from the IFP in $ {\mathfrak{N}_{BB}}_{sm} $.

As in \cite{Abrahao2017published}, we pursue overarching mathematical theorems, so we choose to deal with time-varying directed graphs \cite{Pan2011,Wehmuth2015a}. 
Note that the static case is covered by a particular case of dynamical networks in which the topology does not change over time. 
And the undirected case can be seen as a graph in which each undirected edge (or line) represents two opposing arrows.
As defined in Section~\ref{subsectionGraphsandnetworks}, $ G_t =( \mathrm{V}, \mathscr{E}, \mathrm{T} ) $ are time-varying graphs (TVGs) as in~\cite{Costa2015a,Wehmuth2016b}. 
These are special cases of MAGs which have only one additional aspect relative to variation over time, besides the set of vertices. 

As $ {\mathfrak{N}_{BB}}_{sm} $, the algorithmic networks $  \mathfrak{N}_{SBB} $, which we will define in Definition~\ref{BdefN_SBB}, get their graph topologies from a family of dynamical networks that has a certain diffusion measure as a common feature, in particular, a small diameter compared to the network size (see Definition \ref{BdefTemporaldiameter}).  
In Definition~\ref{BdefFamilyG_sm}, we define $ \mathbb{G}_{sm}(f,t,1) $ as a \emph{family} of unique sized time-varying graphs which shares 
\[ f(i,t,1) = D(G_t,t) = \mathbf{O}\big( \lg( i ) \big) \]
as a common property, where $i$ is the number of nodes and $ D(G_t,t) $ is the temporal diffusion diameter.

Moreover, as $ {\mathfrak{N}_{BB}}_{sm} $, the populations $ \mathfrak{P}_{SBB}( N , s ) $ of nodes/programs in Definition \ref{BdefP_SBB} of $  \mathfrak{N}_{SBB} $ are composed of randomly generated prefix Turing machines (or randomly generated self-delimiting programs) that are represented in a self-delimiting universal programming language $ \mathbf{L_U} $. 
These populations are also synchronous with respect to halting cycles, that is, in the end of a cycle (or communication round, as in distributed computing) every node returns its partial and final outputs at the same time. 
Nodes that do not halt in any cycle always return as final output the lowest fitness/payoff, that is, the integer value $ 0 $.
Here, a straightforward interpretation is that nodes that eventually do not halt in a cycle are ``killed''\footnote{ See also \cite{Abrahao2017published,Chaitin2012,Chaitin2013} for a complete evolutionary formalization of this property. Note that now there is a population of software, while in\cite{Chaitin2012,Chaitin2013} there is only one single organism at the time. }, so that their final output has the ``worst'' fitness/payoff.

Now, unlike the networked population $  {\mathfrak{P}_{BB}}(N)  $ in \cite{Abrahao2017published}, described in Section~\ref{subsectionBBIG}, the networked population $  {\mathfrak{P}_{SBB}}( N , s )  $ follows a modified version of the IFP .
Instead of just returning as final output the largest integer shared by the neighbors, the SIFP ensures that, at the last cycle, every node employs the network input $ w $ together with the largest integer shared by its neighbors to calculate a partial computable function in such a way that every node returns as final output the value of
\[
\mathbf{U} \left( s \circ x \circ w \right) \text{ ,}
\]
where $ x $ was the latest largest integer shared through the network.
Note that this procedure is a \emph{global} information-sharing (or communication) protocol.
In summary, the SIFP makes every node obeys a protocol such that, 
after the first $ c_0 $ node cycles (i.e., after the first rounds of isolated partial outputs), every 
node is obliged to always imitate the neighbor (or itself, if the very node has partially output 
the largest integer in comparison with its neighbors' partial outputs) that has partially output 
the largest integer $x$, repeating this last value as its own partial output in the next node cycle.
Since we are dealing only with synchronous algorithmic networks, these global communication protocols apply at the end of each node cycle (or communication round). 
Finally, the last node cycle is spent in order to cause each node to only return a final output in the form $\mathbf{U} \left( s \circ x \circ w \right) $.
Thus, the SIFP is formally defined as:

\begin{amsdefinitionundersection}\label{BdefSynergisticIFP}
	Let $ s \in \mathbf{L_U} $.
	We say a population $ \mathfrak{P} $ follows a (global) \emph{synergistic imitation-of-the-fittest protocol} (SIFP) for program $ s \in \mathbf{L_U} $ \textit{iff} every \emph{networked} node/program always obeys the procedure: 
	
	\begin{enumerate}[label=\upshape(\Roman*),ref=\theamsdefinitionundersection (\Roman*)] 
		\item \label{procedureIFP} for every $ o_j,o_i \in \mathfrak{ P } $ and $ c,c-1 \in \mathfrak{C} 
		$,
		\begin{enumerate}
			\item \label{clauseFirstcycle1}  if $ \max\{ c \mid c \in \mathfrak{C} \} = 1 $, then
			\[
			p_{o_j,c} = \mathbf{U'}( o_j \circ w )
			\]
			
			\item \label{clauseFirstcycle2} if $c=1$ and $ c \neq \max\{ c \mid c \in \mathfrak{C} \} $, 
			then
			\[
			p_{o_j,c} = w \circ o_j \circ \mathbf{U'}( o_j \circ w )
			\]
			
			\item \label{clauseIFPmain} if $ c \neq 1 $ and $ c \neq \max\{ c \mid c \in \mathfrak{C} \} 
			$, then 
			\[
			p_{o_j,c} = w \circ o_i \circ k 
			\]
			\noindent where 
			\[ k = \max \left\{ x \; \middle| \, \small
			\arraycolsep=1.3pt
			\begin{array}{cl}
			& p_{o_j,c-1}= w \circ o_i \circ x \, \\
			\lor & \, w \circ o_i \circ x \in \left\{ p_{o_i,c-1} \middle| 
			\Small \begin{array}{cl}
				 & o_i \in \mathbf{X}_{neighbors}(o_j, c-1)  \\
				\land  & i \in \mathbb{N}  \\
				\land & c-1 \in \mathfrak{C} 
			\end{array}
			\right\} 
			\end{array}
			\right\} 
			\]
			
			\item \label{clauseLastcycle} if $ c = \max\{ c \mid c \in \mathfrak{C} \} $ and $ p_{o_j,c-1} = w \circ o_i \circ x $, then
			\[
			p_{o_j,c} = \mathbf{U} \left( s \circ x \circ w \right)
			\]
		\end{enumerate}
	\end{enumerate}
\end{amsdefinitionundersection} 

It is important to remark that Definition~\ref{BdefSynergisticIFP} only applies to the networked case. 
If the population $ \mathfrak{ P } $ is isolated, no node can communicate with others. 
Therefore, as we will formalize in Definition~\ref{BdefL_SBB}, no protocol applies in the isolated case. 
Formally:

\begin{amsdefinitionundersection}\label{BdefL_SBB}
	Let $ {\textbf{L}}_{SBB} \subset \mathbf{L_U} $  be a language of programs in the form $ P_{sprot} \circ p \in \mathbf{L_U} $ where $ p \in \mathbf{L_U} $. The prefix $P_{sprot}$ is any program that always ensures that, if the node/program $ P_{sprot} \circ p $ \emph{is networked} and running on $ \mathbf{U'} $, then $P_{sprot} \circ p$ obeys the \emph{synergistic imitation-of-the-fittest protocol} as in Definition~\ref{BdefSynergisticIFP} for some arbitrary program $ s \in \mathbf{L_U} $. Otherwise, if  the node/program $ P_{sprot} \circ p $ is \emph{isolated} and running on $ \mathbf{U'} $, then, for every $ w \in \mathbf{L_U} $, $\mathbf{U'}(P_{sprot} \circ p \circ w ) = \mathbf{U'}(p \circ w)$ and every subsequent node cycle works like a reiteration of partial outputs as immediate respective next inputs for the same program $p$.
\end{amsdefinitionundersection}

Note that the isolated case may be equivalently represented by an algorithmic network built on a 
population of $ p \in \mathbf{L_U} $ that does not follow any information-sharing protocol and 
the topology of the MultiAspect Graph (MAG) is composed by one-step self-loops on each 
node only.

From Definition~\ref{BdefL_SBB}, we can now formalize the population $ \mathfrak{P}_{SBB}( N , s ) $:

\begin{amsdefinitionundersection}\label{BdefP_SBB}
	Let $ \mathfrak{P}_{SBB}( N , s ) $ be the same population  $ \mathfrak{P}_{BB}(N)  $ in \cite{Abrahao2017published}, except for using the language $ \mathbf{L}_{SBB} $ as the support set instead of $ \mathbf{L}_{BB} $.
\end{amsdefinitionundersection}

Thus, we can now formally define the studied model of algorithmic networks $\mathfrak{N}_{SBB} $ as a modification of the algorithmic networks $ {\mathfrak{N}_{BB}}_{sm} $ in \cite{Abrahao2017published}.
In summary, $ \mathfrak{N}_{SBB} $ is a synchronous algorithmic network populated by $N$ randomly generated nodes (i.e., programs) such that, after the first cycle (or arbitrary $c_0$ cycles), it starts a diffusion process of the biggest partial output (given at the end of the first cycle) determined by the network topology of the TVG $ G_t $. 
More specifically:  
before the first cycle each node receives a network input $w$, which is given to every node in the network;
then, before the first network time interval, one or $ c_0 $ cycles are spent in which each node runs separately, repeating its respective first partial output that will be shared; 
from then on, as determined by the SIFP in Definition~\ref{BdefSynergisticIFP}, the plain diffusion of larger integers starts through the respective dynamical network $ G_t \in \mathbb{G}_{sm}(f,t,1)$, so that, at each time interval, the SIFP ensures that a fitter node always ``infects'' its immediate less fit neighbors;
finally, at the last time instant, contagion stops and one cycle (or more) is spent in order to make each node return 
\[
\mathbf{U} \left( s \circ x \circ w \right) \text{ ,}
\]
where $ x $ was the latest largest integer shared by the neighbors at the previous node cycle, as final output.
This way, we formally define:

\begin{amsdefinitionundersection}\label{BdefN_SBB}
	Let $ \mathfrak{N}_{SBB}( N , f , t , 1 , j , s ) $ denote exactly the same algorithmic network $ {\mathfrak{N}_{BB}}_{sm}(N,f,t,1,j) $ in \cite{Abrahao2017published} (see Definition~\ref{BdefN_BB}), except for replacing population $ \mathfrak{P}_{BB}(N) $ with $ \mathfrak{P}_{SBB}( N , s ) $.
\end{amsdefinitionundersection}
%\subsection{Population-size dependent synergistic protocol}

\section{Solving the halting problem through the Busy Beaver imitation game}\label{sectionResults1}

In this section we will prove lemmas, theorems, and corollaries with the purpose of showing that $ \mathfrak{N}_{SBB} $ are algorithmic networks capable of asymptotically solving the halting problem for every network input $ w \in \mathbf{L_U} $ with $ \lg(N) - C_7 \geq \left| w \right|  $, where $ N $ is the network size (i.e., the number of nodes) and $ C_7 $ is a fixed constant that does only depend on the chosen universal programming language.
Moreover, since $ \mathfrak{N}_{SBB} $ are dynamic networks whose functioning is based on the diffusion of the fittest node, we will also show that there is a central node for emergently solving the above halting problem in order that the number of necessary communication rounds are minimized.
In particular, this node is associated with the highest time-reachability centrality among the nodes.

From \cite{Abrahao2017published,Abrahao2018publishedAMS}, it is important to remember that a network Busy Beaver game is a game in which each player is trying to calculate the largest integer it can using the information shared by its neighbors. 
For the present purposes, as in \cite{Abrahao2017published}, the population $ \mathfrak{P}_{SBB}\left( N , s \right) $ in the studied algorithmic networks $ \mathfrak{N}_{SBB}$, which is playing a particular type of network Busy Beaver game, is in fact limited to simple imitation performed by a randomly generated population of programs. 
This Busy Beaver imitation game (BBIG) \cite{Abrahao2017published} is a particular case of the Busy Beaver game in which every node can only propagate the largest integer. 
It configures a simple imitation-of-the-fittest procedure. 
However, unlike \cite{Abrahao2017published,Abrahao2018publishedAMS}, the last node cycle is devoted to employ this diffusion of the fittest to solve a problem that can be \emph{partially} computed by program $ s $. 
Although partial or final outputs are always defined in algorithmic networks $ \mathfrak{N}_{SBB}$ due to the oracle-sensitiveness property \cite{Abrahao2017published} of the population $ \mathfrak{P}_{SBB}\left( N , s \right) $, these outputs may not match every function value for every input in some cases. 
For example, our central results in Theorem~\ref{BthmHaltingproblem} demands a restriction on the domain of possible network inputs in order that every input in this domain generates the correct function value.
Nevertheless, in the limit when the population grows indefinitely, one can say that an infinite family of algorithmic networks $ \mathfrak{N}_{SBB}$ make every node asymptotically compute a total function (in the case, the very characteristic function of the halting problem).

While algorithmic networks $ \mathfrak{N}_{BB} $ in \cite{Abrahao2017published} can be seen as playing an optimization procedure where the whole pursues the increase of the average fitness/payoff through diffusing on the network the best randomly generated solution, these algorithmic networks $ \mathfrak{N}_{SBB} $ can be seen as playing an optimization procedure where the whole pursues the increase of each node's capability of solving a common problem through diffusing on the network the best randomly generated solution in the smallest number of communication rounds as possible \cite{Abrahao2017published}.
Thus, the nodes in algorithmic networks $ \mathfrak{N}_{BB} $ may be seen as competing with each other, as in multi-agent 
systems from a game-theoretical approach \cite{Abrahao2017published,Abrahao2018publishedAMS,Abrahao2016b};
on the other hand, the present model $ \mathfrak{N}_{SBB} $ may be seen as nodes/programs computing using network's 
shared information to solve a common purpose, as the classical approach in distributed 
computing.
In this sense, this addition of perspective in such models of algorithmic networks is bridging a competition or individualistic-centered view of emergence to a synergistic-centered view of emergence.
In Section~\ref{sectionSynergy} we will define and explore such synergy in algorithmic networks.

First, we define a total computable function that is capable of deciding whether a program halts or not if the respective large enough computation time is informed as input.

\begin{amsdefinitionundersection}\label{BdefFunctionhalt}
	Let $ \mathrm{p}_{halt} \in \mathbf{L_U} $ be a program of $ \mathbf{U} $ that computes a \emph{total} recursive function such that
	\[
	\begin{array}{rl}
		\mathbf{U}\left( \mathrm{p}_{halt} \circ n \circ p \right) 
		\; = &
		\textnormal{the output of the algorithm that only lets } \\
		 &
		\textnormal{$ \mathbf{U} $ runs with $ p \in \mathbf{L_U}$ as its input} \\
		 &
		\textnormal{for $ t \leq n + 1 $ computation time until } \\
		 & \textnormal{it halts or not and, then,} \\
		  & \textnormal{returns }
		\begin{cases}
			\left( `` \textit{non-halting} " \right)_{ \mathbf{L_U} } 
			& \textnormal{, if } t > n \in \mathbb{N} \\
			\left( `` \textit{halting} " \right)_{ \mathbf{L_U} }
			& \textnormal{, otherwise}
		\end{cases}
	\end{array}
	\]
\end{amsdefinitionundersection}

Note that $ \mathrm{p}_{halt} $ in Definition~\ref{BdefFunctionhalt} always computes a total function for every $ n \in \mathbb{N}  $ and every $ p \in \mathbf{L_U} $ because $ \mathbf{U} $ can either halt or not halt on any program $p$ in $ n $ computation time. 
As a consequence:

\begin{amslemmaundersection}\label{BlemmaFunctionp_halt}
	Let $ N \geq \left|  \mathrm{p}_{+1} \circ \mathrm{p}_{T} \circ p \right| \in \mathbb{N} $, where $ p \in \mathbf{L_U} $.
	Then, for every $ x \geq BB(N)  $,
	\[
	\mathbf{U}\left( \mathrm{p}_{halt} \circ x \circ p \right) 
	=
	\begin{cases}
		\left( `` \textit{non-halting} " \right)_{ \mathbf{L_U} } 
		& \textnormal{, if $ \mathbf{U} $ does not halt on $ p $ } \\
		\left( `` \textit{halting} " \right)_{ \mathbf{L_U} }
		& \textnormal{, if $ \mathbf{U} $ halts on $ p $}
	\end{cases}
	\]
	
	\begin{proof}
		The proof follows directly from Definition~\ref{BdefFunctionhalt}.
		Since $ \mathbf{U} $ can only halt or, exclusively,  not halt on $ p $, we divide the proof in two cases:
		\begin{enumerate}
			\item  If the universal machine $ \mathbf{U} $ halts on $ p $, then the value $ T( \mathbf{U} , p ) $ will be well defined.
				Hence, the value $ \mathbf{U}\left( \mathrm{p}_{+1} \circ \mathrm{p}_{T} \circ p  \right) $ will also be well defined.
				Therefore, from Definition~\ref{BdefFunctionBB}, we will have that $ BB(N) \geq  \left(\mathbf{U}\left( \mathrm{p}_{+1} \circ \mathrm{p}_{T} \circ p  \right) \right)_{10}  > \left( \mathbf{U}\left( \mathrm{p}_T \circ p \right) \right)_{10}  = T( \mathbf{U} , p ) $.
				Then, 
				\begin{equation*}
				\begin{aligned}
					x \geq BB(N)
					& \implies
					x > T( \mathbf{U} , p ) 
					\implies \\
					& \implies
					\mathbf{U}\left( \mathrm{p}_{halt} \circ x \circ p \right)  = \left( `` \textit{halting} " \right)_{ \mathbf{L_U} } 
					\text{ ;}
				\end{aligned}
				\end{equation*}
				
			\item If the universal machine $ \mathbf{U} $ does \emph{not} halt on $ p $, then the value $ T( \mathbf{U} , p ) $ will not be well defined.
				Therefore, for every $ t \leq x $, we will have $ T( \mathbf{U} , p ) \neq t $.
				Then, $ \mathbf{U}\left( \mathrm{p}_{halt} \circ x \circ p \right)  = \left( `` \textit{non-halting} " \right)_{ \mathbf{L_U} } $.
		\end{enumerate}
	\end{proof}
\end{amslemmaundersection}

Now, from \cite{Abrahao2017published}, we translate its first lemma to the new algorithmic network model $  \mathfrak{N}_{SBB} $, showing how to harness the implications of the law of large numbers in a program-size probability distribution \cite{Abrahao2017published}:

\begin{amslemmaundersection}\label{BlemmaA_maxinN_halt}
	Let $  \mathfrak{N}_{SBB}( N , f , t , 1 , j , s )  = ( G_t, \mathfrak{P}_{SBB}( N , s ) , b_j ) $ be an algorithmic network as in Definition~\ref{BdefN_SBB}.
	Then, with probability arbitrarily close to 1 as $N$ increases toward infinity, we will have that there are constants $ C_{BB} $ and $ C_4 $ such that
	\[ A(x) \geq \lg(N) - C_4 \text{ ,} \]
	\noindent where 
	\[ x = \max \left\{ y \, \middle\vert \, \begin{array}{lc} p_{o_i,c-1} = w \circ o_k \circ y & \land \\ o_i, o_k \in \mathfrak{P}_{SBB}( N , s ) & \land \\ c = \max\{ c \mid c \in \mathfrak{C} \} \end{array}  \right\} \text{ ,}\]
	$ \max\{ c \mid c \in \mathfrak{C} \} \geq 2 $, and $ w $ is the network input.
	\noindent In addition, for every $ o_i, o_k \in \mathfrak{P}_{SBB}( N , s ) $ with $ p_{o_i,c-1} = w \circ o_k \circ x $, $ c \geq c_0 + t + f( N, t, 1 ) + 1  $, and $ c = \max\{ c \mid c \in \mathfrak{C} \}  $, we will have
	\[
	x \geq BB( \lg(N) - C_{BB} )
	\]
	
	\begin{proof}
		This proof of $ A(x) \geq \lg(N) - C_4 $ is totally analogous to the proof of Lemma 5.1 in \cite{Abrahao2017published}.
		Just note that, from Definition~\ref{BdefN_SBB} and \cite{Abrahao2017published}, we also have that  $ {\mathfrak{N}_{BB}}_{sm}(N,f,t,1,j) $ denotes the same algorithmic network $ \mathfrak{N}_{BB}(N,f,t,\tau,j) $ in \cite{Abrahao2017published}, except for replacing family $ \mathbb{G}(f, t, \tau) $ with family $ \mathbb{G}_{sm}(f,t,1) $.
		In addition, from Definition~\ref{BdefSynergisticIFP} and the definition of $ A_{max} $ in \cite{Abrahao2017published}, we have that $ A(x) = A_{max} $, which follows from the fact that the IFP in \cite{Abrahao2017published} is only different from the SIFP in just ensuring that nodes returns $ x $ instead of $ \mathbf{U}\left( s \circ x \circ w \right) $ in the last cycle as the final output.
		To show the second part of the Lemma~\ref{BlemmaA_maxinN_halt} that $ x \geq BB( \lg(N) - C_{BB} ) $, it suffices to note that, from Definition~\ref{BdefSynergisticIFP}, every networked node only imitates the fittest neighbor after the first node cycle. 
		Thus, since $ c \geq c_0 + t + f( N, t, 1 ) + 1  $,  we have from Definition~\ref{BdefFamilyG_sm} that the number of node cycles will cover the temporal diffusion diameter and it will be enough to make any fittest first partial output, which is at least as fit as $ BB( \lg(N) - C_{BB} ) $, propagate to every other node.
	\end{proof}
\end{amslemmaundersection}

Thus, we can now combine the previous results to build a new theorem.
Theorem~\ref{BthmHaltingproblem} basically asymptotically assures that, if enough communication rounds are expended, matching the network temporal diffusion diameter (which is small compared to the population/network size), then every node is expected to solve the halting problem for program with length dominated by a logarithmic order of the population/network size.
Therefore, increasing the population/network size can make such algorithmic networks emergently solve a increasing number of instances of the halting problem such that, in the limit when the population grows indefinitely, all instances of the halting problem will be statistically covered.

\begin{thmundersection}\label{BthmHaltingproblem}
	Let $  \mathfrak{N}_{SBB}( N , f , t , 1 , j , s )  = ( G_t, \mathfrak{P}_{SBB}( N , s ) , b_j ) $ be an algorithmic network as in Definition~\ref{BdefN_SBB} such that $  \mathfrak{N}_{SBB}( N , f , t_{z_0} , 1 , j , \mathrm{p}_{halt} ) $ is well defined. 
	Let 
	\[ \myfunc{c \,}{ \mathbb{N} } { \mathfrak{C_{BB}} } { x } { c(x)=y } \]
	be a non-decreasing total computable function such that 
	\[ c(z_0 + f( N, t_{z_0}, 1 ) + 2) \geq c_0 + z_0 + f( N, t_{z_0}, 1 ) + 2 \text{ ,} \] where $ t_{z_0} \in \mathrm{T}(G_t) $.
	Then, there is a constant $ C_7 $ such that, for large enough $ N $ and for every network input $ w \in \mathbf{L_U} $ with $ \lg(N) - C_7 \geq \left| w \right|  $, we will have that every node in $  \mathfrak{N}_{SBB}( N , f , t_{z_0} , 1 , j , \mathrm{p}_{halt} ) $ decides whether $ \mathbf{U} $ halts or not on $ w $ in  $ c(z_0 + f( N, t_{z_0}, 1 ) + 2) $ node cycles with probability arbitrarily close to $1$.
	%In addition, for large enough $ N $, there is at least one central node that decides whether $ \mathbf{U} $ halts or not on $ w $ with probability arbitrarily close to $1$.
	%Then, for every non-decreasing total computable function $ \myfunc{c}{ \mathbb{N} } { \mathfrak{C_{BB}} } { x } { c(x)=y } $, where $ t_{z_0} \in \mathrm{T}(G_t) $ and $ c(z_0 + f( N, t_{z_0}, 1 ) + 2) \geq c_0 + z_0 + f( N, t_{z_0}, 1 ) + 2 $, we will have that every node in $  \mathfrak{N}_{SBB}( N , f , t_{z_0} , 1 , j , \mathrm{p}_{halt} ) $ decide whether $ \mathbf{U} $ halts or not on $ w $ in  $ c(z_0 + f( N, t_{z_0}, 1 ) + 2) $ node cycles.

	\begin{proof}
		This proof follows from combining Lemma~\ref{BlemmaA_maxinN_halt} with Lemma~\ref{BlemmaFunctionp_halt}.
		We have from Clause~\ref{clauseLastcycle} in Definition~\ref{BdefSynergisticIFP} that, for every $ o_i \in \mathfrak{P}_{SBB}( N , \mathrm{p}_{halt}  ) $, 
		\[ p_{ o_i , c' } = \mathbf{U} \left( \mathrm{p}_{halt} \circ x \circ w \right) \text{ ,} \]
		\noindent where $ c' = c(z_0 + f( N, t_{z_0}, 1 ) + 2) $.
		In addition, we have from Lemma~\ref{BlemmaA_maxinN_halt} that $ x \geq  BB( \lg(N) - C_{BB} ) $ holds with probability arbitrarily close to $1$ as the population size $N$ tends to infinity.
		Let $ C_7 $ be a constant such that 
		\[  \left| w \right| + C_7  \geq \left|  \mathrm{p}_{+1} \circ \mathrm{p}_{T} \circ w \right| + C_{BB} \text{ .} \]
		Therefore, for every network input $ w \in \mathbf{L_U} $ with $ \lg(N) - C_7 \geq \left| w \right|  $, we will have that $  \lg(N)  \geq \left|  \mathrm{p}_{+1} \circ \mathrm{p}_{T} \circ w \right| + C_{BB} $ and, hence,
		from Lemma~\ref{BlemmaFunctionp_halt}, that
		\[  
		p_{ o_i , c' }
		 =
		 \begin{cases}
		 \left( `` \textit{non-halting} " \right)_{ \mathbf{L_U} } 
		 & \textnormal{, if $ \mathbf{U} $ does not halt on $ w $ } \\
		 \left( `` \textit{halting} " \right)_{ \mathbf{L_U} }
		 & \textnormal{, if $ \mathbf{U} $ halts on $ w $}
		 \end{cases}
		 \]
	\end{proof}
\end{thmundersection}

Theorem~\ref{BthmHaltingproblem} looks at all nodes in the algorithmic network.
However, one may extract from this result the presence of privileged nodes in solving the respective halting problem.
To this end, we first define a general node centrality for distributed processing:

\begin{amsdefinitionundersection}\label{BdefTimereachablecentralvertexinsolvingaproblem}
	Let $ \mathfrak{N} = (G, \mathfrak{P}, b) $ be a well-defined algorithmic network in the language $ \mathbf{L_U} $ and machine $ \mathbf{U'} $.
	We define a \emph{central node} $ o_{cen} $ that can compute a function 
	\[ \myfunc{f \,}{ X \subseteq \mathbf{L_U} }{ \mathbf{L_U} }{ w }{ f(w) } \]
	\noindent in the minimum number $ c_{min} $ of node cycles when networked and that does not compute this function when isolated in $ c_{min} $ node cycles \textit{iff} for every $ w \in X $, the final output $ p_{o_{cen},\max\{ c \mid c \in \mathfrak{C}( o_{cen} ) \}  }  $ of the networked node $ o_{cen} $ hold as
	\[ p_{o_{cen}, c_{min}=\max\{ c \mid c \in \mathfrak{C}( o_{cen} ) \}  } = f(w) \]
	\noindent such that, for every $ c' < c_{min} $,
	\[
	\mathbf{U}\left(  p_{iso} ( o_{cen} , c' , w )  \right) \neq f(w)
	\]
	and, for every $ o \in \mathfrak{P}  $, if
	\[
	\mathbf{U}\left(  p_{iso} ( o , c' , w )  \right) \neq f(w)
	\]
	and
	\[ p_{o, c =\max\{ c \mid c \in \mathfrak{C}( o ) \} } = f(w) \text{ ,}\]
	then $ c_{min} \leq c $. 
\end{amsdefinitionundersection}

Now, we come back to node centralities in network science in order to rank the node that can be quickly accessible by an arbitrary diffusion from an arbitrary fraction of the nodes. From \cite{Costa2015a}:

\begin{amsdefinitionundersection}\label{BdefInvertedd_t}
	Let $\overline{d_t}(G_t, t_i, u, \tau)$ be the minimum number of time intervals (non-spatial steps or, specially in the present article, node cycles) for a diffusion starting on any vertex $ v \in X $ of a fraction $ \tau = \left| X \right| $ of vertices in the TVG $G_t$ at time instant $t_i$ to reach vertice $u$, where $ X \in \mathcal{P}\left( \mathrm{V}\left( G_t \right) \right) $ is arbitrary.
\end{amsdefinitionundersection}

In this sense, if one consider any possible fraction of nodes at the same time, we can define a node centrality based on the temporal diffusion diameter \cite{Abrahao2017published}:

\begin{amsdefinitionundersection}\label{BdefTimereachablecentralvertex}
	Let $ G_t $ be a TVG with $ D(G_t,t) \neq \infty $ and $ \left| \mathrm{V}\left( G_t \right) \right| \geq 2 $.
	We define the \emph{time-reachability centrality} of a vertex $u$ in the TVG $ G_t $ from time instant $ t \in \mathrm{T}( G_t ) $ as 
	\[
	\frac{1}{\overline{d_t}(G_t, t, u, 1)}
	\]
\end{amsdefinitionundersection}

In addition: 

\begin{amsdefinitionunderamsdefinitionundersection}\label{BdefTimereachablenodes}
	We define the set of the
	%most \emph{time-reachable} (i.e., the set of vertices with the highest time-reachability centrality) 
	vertices with time-reachability centrality $ x \leq D(G_t,t) \neq \infty $ in a TVG $ G_t $ from time instant $ t \in \mathrm{T}( G_t ) $ as
	\[
	\mathbf{X}_{treach}( G_t , t , x )
	=
	\left\{ u \, \middle| \, x = \frac{1}{\overline{d_t}(G_t, t, u, 1)} \land u \in \mathrm{V}( G_t ) \right\}
	\]
\end{amsdefinitionunderamsdefinitionundersection}

Note that the condition $ D(G_t,t) \neq \infty $ immediately assures that Definitions~\ref{BdefTimereachablecentralvertex} and~\ref{BdefTimereachablenodes} are well defined.

Finally, Theorem~\ref{BthmHaltingproblem} implies that the node centrality for distributed processing and the node centrality for complex network's dynamics can be combined to find a node that can only solve the halting problem when networked in the least amount of communication rounds (i.e., node cycles):

\begin{amscorollaryunderthmundersection}\label{BcorCentralnodeHalting}
	Let $  \mathfrak{N}_{SBB}( N , f , t , 1 , j , s )  = ( G_t, \mathfrak{P}_{SBB}( N , s ) , b_j ) $ be an algorithmic network as in Definition~\ref{BdefN_SBB} such that $  \mathfrak{N}_{SBB}( N , f , t_{z_0} , 1 , j , \mathrm{p}_{halt} ) $ is well defined. 
	Let $ \myfunc{c}{ \mathbb{N} } { \mathfrak{C_{BB}} } { x } { c(x)=y } $ be a non-decreasing total computable function with $ t_{z_0} \in \mathrm{T}(G_t) $ and $ c(z_0 + f( N, t_{z_0}, 1 ) + 2) \geq c_0 + z_0 + f( N, t_{z_0}, 1 ) + 2 $.
	Then, there is a constant $ C_7 $ such that, for large enough $ N $ and for every network input $ w \in \mathbf{L_U} $ with $ \lg(N) - C_7 \geq \left| w \right|  $, there is at least one central node (as in Definition~\ref{BdefTimereachablecentralvertexinsolvingaproblem}) with the respective highest time-reachability centrality (as in Definition~\ref{BdefTimereachablecentralvertex}) in the algorithmic network  $  \mathfrak{N}_{SBB}( N , f , t_{z_0} , 1 , j , \mathrm{p}_{halt} ) $ that decides whether $ \mathbf{U} $ halts or not on $ w $ in $ \mathbf{O}\left( \lg\left( N \right) \right) $ node cycles with probability arbitrarily close to $1$.

	\begin{proof}
		This proof follows from Theorem~\ref{BthmHaltingproblem}.
		Since $ G_t \in \mathbb{G}_{sm}(f,t_{z_0},1) $, we will have that Definition~\ref{BdefTimereachablecentralvertex} is well defined for every vertex in the TVG $ G_t $.
		Now, we take a vertex $ v \in \mathrm{V}( G_t ) $ with the highest time-reachability centrality as in Definition~\ref{BdefTimereachablecentralvertex} such that,  for every $ c'' < c' $,
		\[
		\mathbf{U}\left(  p_{iso} ( o_{cen} , c'' , w )  \right) \neq f(w) \text{ ,}
		\]
		where $ c' =  c(z_0 + f( N, t_{z_0}, 1 ) + 2) $ and $  b_j(v,t_{c-1})=( o_{cen} , c_0 + c ) $.
		% Hence, we have that there is at least one most time-reachable vertex $ v \in \mathbf{X}_{treach}( G_t , t_{z_0} , x ) $.
		Thus, we trim the necessary latest time instants in the set of time instant $ \mathrm{T}( G_t ) $ in order that one can define another TVG $ G'_t =( \mathrm{V}( G_t ), \mathscr{E'}, \mathrm{T'} ) $ such that 
		\[ 
		\mathrm{T'}( G'_t ) = \mathrm{T}( G_t ) \setminus \left\{ t \, \middle| \, t > t_{z_0} +  \overline{d_t}(G_t, t_{z_0} , v, 1) \right\} 
		\]
		\noindent and, for every $ e \in \mathscr{E'}( G'_t ) $, one has $ e \in \mathscr{E}( G_t ) $.
		Then, we replace $ G_t $ in $ ( G_t, \mathfrak{P}_{SBB}( N , s ) , b_j ) $ with $ G'_t $.
		Note that, from Definition~\ref{BdefFamilyG_sm}, we have that 
		\[ \overline{d_t}(G_t, t_{z_0} , v, 1)  \leq D( G_t , t_{z_0} )  = f( N, t_{z_0}, 1 ) = \mathbf{O}\left( \lg\left( N \right) \right) \text{ .} \]
		Now, we take any non-decreasing total computable function $ \myfunc{c}{ \mathbb{N} } { \mathfrak{C_{BB}} } { x } { c(x)=y } $ such that $ t_{z_0} \in \mathrm{T}(G_t) $ and 
		\[ c_0 + \mathbf{O}\left( \lg\left( N \right) \right) +1 = c(z_0 + f( N, t_{z_0}, 1 ) + 2) \geq c_0 + z_0 + f( N, t_{z_0}, 1 ) + 2 \text{ .}\]
		Therefore, from Definition~\ref{BdefN_SBB}, there is a correspondent node $ o_{cen} \in \mathfrak{P}_{SBB}( N , \mathrm{p}_{halt} ) $ that assumes position of the vertex $ v \in \mathrm{V}( G_t ) = \mathrm{V}( G'_t )$ such that, from Theorem~\ref{BthmHaltingproblem},
		\begin{equation*}
		\begin{aligned}
			p_{ o_{cen}, c_0 + \mathbf{O}\left( \lg\left( N \right) \right) +1 } 
%			&= 
%			p_{o_i,c(z_0 + f( N, t_{z_0}, 1 ) + 2)} = \\
			&=
			\begin{cases}
			\left( `` \textit{non-halting} " \right)_{ \mathbf{L_U} } 
			& \textnormal{, if $ \mathbf{U} $ does not halt on $ w $ } \\
			\left( `` \textit{halting} " \right)_{ \mathbf{L_U} }
			& \textnormal{, if $ \mathbf{U} $ halts on $ w $}
			\end{cases}
		\end{aligned}
		\end{equation*}
%		\[\displaystyle{l_i = \frac{1}{n} \sum_j d_{i,j}}\]
%		\[\displaystyle{C_i = \frac{1}{l_i} = \frac{n}{\sum_j d_{i,j}}} \]
	\end{proof}
\end{amscorollaryunderthmundersection}

%\subsection{Solving the halting time problem through the Busy Beaver imitation game}\label{sectionResults2}

\section{Algorithmic synergy}\label{sectionSynergy}

Following the same spirit from the emergent algorithmic complexity of a node introduced in \cite{Abrahao2017published,Abrahao2018publishedAMS}, another interesting topic is whether algorithmic networks and algorithmic information theory are sufficient to deal with the problem of measuring synergistic information
\cite{Lizier2018,Griffith2014a}
%\cite{Williams2010, Griffith2012, Griffith2014, Oizumi2014, Wibral2015, Quax2017} 
or not.
This problem is usually stated within the context of multivariate information theory for stochastic dynamical systems.
Generally speaking, it concerns measuring the amount of information in an arbitrary collection of random variables $ X_1 , \dots , X_n $ that predicts another random variable $ Y $, but that it is not contained in (or does not derive from) any individual random variable $ X_i $, where $ 1 \leq i \leq n $, or from combinations of proper subsets of the set $ \left\{ X_1 , \dots , X_n \right\} $, which is given by the partial information diagrams (i.e., PI-diagrams) \cite{Griffith2014a}.

On the other hand, from \cite{Abrahao2017published,Abrahao2018publishedAMS}, note that emergent algorithmic complexity directly gives a formal measure of irreducible information \cite{Li1997, Calude2002, Chaitin2004,Calude2009,Grunwald2008} that emerges when comparing the networked case with the isolated case.
Thus, if one assumes the definition of synergy as the general phenomenon in which the whole system is irreducibly better in solving a common problem than the ``sum'' (or the ``union'') of its parts taken separately, as the problem described in the previous paragraph, then there should be an immediate extension of emergent algorithmic complexity to algorithmic synergistic information.

To tackle this problem, we introduce in this section a formalization of one type of algorithmic synergistic information in the context of algorithmic networks.
Thus, instead of studying synergy in stochastic processes, we will be studying synergy in deterministic systems, in particular, in networks of computable systems.
That is, we are focusing on the general problem of measuring the amount of algorithmic information in an arbitrary collection of nodes necessary to calculate a function, but that could not be performed by any individual isolated node or by any combination of proper subnetworks of the entire network.
In particular, we start by formalizing a measure of average algorithmic synergistic information for individual nodes when comparing the fully networked case with the isolated case.
Thus, we leave the joint cases and subnetwork cases for future research.

\begin{amsdefinitionundersection}\label{BdefLocalsynergy}
	Let $ \mathfrak{N} = (G, \mathfrak{P}, b) $ be a well-defined algorithmic network.
	We define the \emph{local algorithmic synergy} of a node $ o_i \in \mathfrak{P} $ toward function
	\[
	\myfunc{f \,}{ X \subseteq L }{ L }{ x }{ f(x) }
	\]
	\noindent in $ c $ node cycles with network input $w$ as
	\begin{equation*}
	\begin{aligned}[center]
	{ \displaystyle{ \myDelta_{ iso }^{ net(b) } } } I_K ( ( o_i , c , w ) : f(w) ) = \\
	I_K ( \mathbf{U}( p_{net}^{b} ( o_i , c , w ) ) : f(w) ) - 
	I_K ( \mathbf{U}( p_{iso} ( o_i , c , w ) ) : f(w) )
	\pm 
	\mathbf{O}(1)
	\end{aligned}
	\text{ ,}
	\end{equation*}
	\noindent where:
	\begin{enumerate}
		\item $ f_o( o_i ) \in L $;
		
		\item $ p_{net}^{b} ( o_i , c , w ) $ represents the program that returns the final output of $o_i$ when networked assuming the position $ v $, where $ b(v,\mathbf{\bar{x}}) = (o_i, b_{ dim( Y ) - 1 }( \mathbf{ \overline{x} } ) ) $, in the MAG $G$ in the specified number of node cycles $c$ with network input $w$; 
		
		\item $ p_{iso} ( o_i , c , w )  $ represents the program that returns the final output of $o_i$ when isolated in the specified number of node cycles $ c $ with network input $w$;
	\end{enumerate}
\end{amsdefinitionundersection}

The reader may find tempting to employ $ I_A( x \, ; y ) $ instead of $ I_K( x \, : y ) $ in Definition~\ref{BdefLocalsynergy} due to the fact that $ I_A( x \, ; y ) $ is invertible and $ I_K( x \, : y ) $ is not (see Definition~\ref{BdefAlgComp} and \cite{Li1997,Downey2010,Chaitin2004}).
In this sense, note that, since $ I_K( x : y ) = A(y) - A( y \, | x ) $, we will have that 
\begin{equation}\label{BequationSimplifiedlocalsynergy}
\begin{aligned}[center]
{ \displaystyle{ \myDelta_{ iso }^{ net(b) } } } I_K ( ( o_i , c , w ) : f(w) ) = \\
A ( f(w) \, | \,\mathbf{U}(  p_{iso} ( o_i , c , w ) )  ) -
A ( f(w) \, | \, \mathbf{U}(  p_{net}^{b} ( o_i , c , w ) ) ) 
\pm 
\mathbf{O}(1)
\end{aligned}
\end{equation}
However, besides the outputs processed by the algorithmic network be in the form $ \mathbf{U}(  p ) $ and not $ {\mathbf{U}(  p ) }^* $, the non-invertibility of $ I_K( x \, : y ) $ actually captures the notion of towardness in computing the function $f$.
For example, the algorithmic network may be emergently generating the necessary information to compute function $f$ at the same time that function $f$ does not give the necessary information to determine the emergent behavior of the algorithmic network.
This way, the non-invertibility would be a sound property.
In fact, investigating the cases in which the measure in Definition~\ref{BdefLocalsynergy} is invertible is an interesting future research.

Moreover, a constant represented by $ \mathbf{O}(1) $ in the Definition~\ref{BdefLocalsynergy} is employed in order to deal with some non intuitively correct cases that may appear depending on the chosen universal programming language.
For example, when
\[
A ( f(w) \, | \,\mathbf{U}(  p_{iso} ( o_i , c , w ) )  )
%=
%A ( f(w) \, | \, w_{min} )
<
A ( f(w) \, | \, \mathbf{U}(  p_{net}^{b} ( o_i , c , w ) ) ) 
=
A ( f(w) \, | \, f(w) ) 
\]
with $ \mathbf{U}(  p_{iso} ( o_i , c , w ) )  \neq f(w) $
or when
\[
A ( f(w) \, | \,\mathbf{U}(  p_{iso} ( o_i , c , w ) )  )
=
A ( f(w) \, | \, f(w) )
>
A ( f(w) \, | \, \mathbf{U}(  p_{net}^{b} ( o_i , c , w ) ) ) 
\]
with $ \mathbf{U}(  p_{net}^{b} ( o_i , c , w ) )  \neq f(w) $.
Thus, there may be ``blurred intervals'' with respect to algorithmic synergy, so that one cannot decide whether there is a positive value of local algorithmic synergy or not.
In fact, as odd as it may seem, it is in consonance with the equalities and inequalities in algorithmic information theory that hold, except for a constant that only depends on the chosen universal programming language. 
Note that these complexity/information oscillations are expected to happen in algorithmic information theory.

In order to specify the type of algorithmic network from which one is calculating the local algorithmic synergy, we may also denote $ { \displaystyle{ \myDelta_{ iso }^{ net(b) } } } I_K ( ( o_i , c , w ) : f(w) ) $ by
\[
{ \displaystyle{ \myDelta_{ iso }^{ \mathfrak{N } } } } I_K ( ( o_i , c , w ) : f(w) ) 
\]
\noindent or
\[
{ \displaystyle{ \myDelta_{ iso }^{ (G, \mathfrak{P}, b)  } } } I_K ( ( o_i , c , w ) : f(w) )  \text{ .}
\]
\noindent Thus, for the current studied model:
\begin{amsdefinitionunderamsdefinitionundersection} \label{BdefLocalsynergySBB}
	We denote the \emph{local algorithmic synergy} of a node $ o_i \in \mathfrak{P}_{SBB}( N , s )  $ in an algorithmic network $ \mathfrak{N}_{SBB}( N , f , t , 1 , j , s )   $ toward function
	\[
	\myfunc{f \,}{ X \subseteq \mathbf{L_U} }{ \mathbf{L_U} }{ x }{ f(x) }
	\]
	\noindent in $ c $ node cycles with network input $w$ as
	\begin{equation*}
	\begin{aligned}[center]
	{ \displaystyle{ \myDelta_{ iso }^{ \mathfrak{N}_{SBB}( N , f , t , 1 , j , s )  } } } I_K ( ( o_i , c , w ) : f(w) ) = \\
	I_K ( \mathbf{U}( p_{net}^{b_j} ( o_i , c , w ) ) : f(w) ) - 
	I_K ( \mathbf{U}( p_{iso} ( o_i , c , w ) ) : f(w) )
	\pm 
	\mathbf{O}(1)
	\end{aligned}
	\end{equation*}
	\noindent where:
	\begin{enumerate}
		\item $ f_o( o_i ) = P_{sprot} \circ p_i \in \mathbf{L}_{SBB} $ ;
		
		\item $ p_{net}^{b_j} ( o_i ,  c , w ) $ represents the program that returns the final output of $o_i$ \emph{when networked} assuming the position $ v $, where $ b_j(v)=(o_i) $, in the TVG $G_t$ in $c$ node cycles with network input $w$; 
		
		\item $ p_{iso} ( p_i ,  c , w ) $ represents the program that returns the final output of $p_i$ \emph{when isolated} in $c$ node cycles with network input $w$; 
	\end{enumerate}

\end{amsdefinitionunderamsdefinitionundersection}

Furthermore, for a fixed function $ b $, one can define the average value of local algorithmic synergy:

\begin{amsdefinitionundersection}\label{BdefAveragelocalsynergy}
	Let $ \mathfrak{N} = (G, \mathfrak{P}, b) $ be a well-defined algorithmic network.
	We define the \emph{average local algorithmic synergy} of a node $ o_i \in \mathfrak{P} $ toward function
	\[
	\myfunc{f \,}{ X \subseteq L }{ L }{ x }{ f(x) }
	\]
	\noindent in $ c $ node cycles with network input $w$ as
	\begin{equation*}
	\begin{aligned}[center]
	\frac{ \sum\limits_{ o_i \in \mathfrak{P} } { \displaystyle{ \myDelta_{ iso }^{ \mathfrak{N} } } } I_K ( ( o_i , c , w ) : f(w) ) }
	{ \left| \mathfrak{P} \right| }
	\end{aligned}
	\end{equation*}
\end{amsdefinitionundersection}

Then, since population $ \mathfrak{P}_{SBB}( N , s )  $ is randomly generated, one can define the expected local value of the algorithmic synergy of a node for a fixed function $ b_j $ in the current studied case:

\begin{amsdefinitionunderamsdefinitionundersection}\label{BdefExpectedlocalsynergySBB}
	We define the \emph{expected local algorithmic synergy} of a node $ o_i \in \mathfrak{P}_{SBB}( N , s )  $ toward function
	\[
	\myfunc{f \,}{ X \subseteq L }{ L }{ x }{ f(x) }
	\]
	\noindent in $ c $ node cycles (or communication rounds) with network input $w$ as
	\begin{equation*}
	\begin{aligned}[center]
	\mathbf{E}_{  \mathfrak{N}_{SBB} } \left( { \displaystyle{ \myDelta_{ iso }^{ \mathfrak{N}_{SBB}( N , f , t , 1 , j ,s  ) } } } I_K ( ( o_i , c , w ) : f(w) ) \right)
	= \\
	\frac{ \sum\limits_{ o_i \in  \mathfrak{P}_{SBB}( N , s ) } { \displaystyle{ \myDelta_{ iso }^{ \mathfrak{N}_{SBB}( N , f , t , 1 , j , s )  } } } I_K ( ( o_i , c , w ) : f(w) ) }
	{ N }
	\end{aligned}
	\end{equation*}
\end{amsdefinitionunderamsdefinitionundersection}

Now, we can combine the results from Section~\ref{sectionResults1} with the definition of expected local algorithmic synergy in order to make it as large as one may want:

\begin{thmundersection}\label{BthmLocalsynergy}
	Let 
	\[
%	\myfunc{f_h \,}{ X \subseteq \mathbf{L_U} }{ \{ h , \overline{ h }\} \subset \mathbf{L_U} }{ w }
%	{ f_h (w) 
%	=
%	=
%	\begin{cases}
%	\left( `` \textit{non-halting} " \right)_{ \mathbf{L_U} } 
%	& \textnormal{, if $ \mathbf{U} $ does not halt on $ p $ } \\
%	\left( `` \textit{halting} " \right)_{ \mathbf{L_U} }
%	& \textnormal{, if $ \mathbf{U} $ halts on $ p $}
%	\end{cases} 
	\arraycolsep=0.8pt
	\begin{array}{cccl}
	f_h \; : & X \subseteq \mathbf{L_U} & \to & \{ h , \overline{ h }\} \subset \mathbf{L_U} \\
	 & x & \mapsto & f_h(x) 
	=
	 \begin{cases}
	 \overline{ h }= \left( `` \textit{\small non-halting} " \right)_{ \mathbf{L_U} } 
	 & \textnormal{\small , if $ \mathbf{U} $ does not halt on $ x $ } \\
	 h = \left( `` \textit{\small halting} " \right)_{ \mathbf{L_U} }
	 & \textnormal{\small , if $ \mathbf{U} $ halts on $ x $}
	 \end{cases}
	\end{array}
	\]
	\noindent be a function defined on arbitrary $  h \in \mathbf{L_U} $. 
	Let $  \mathfrak{N}_{SBB}( N , f , t , 1 , j , s )  = ( G_t, \mathfrak{P}_{SBB}( N , s ) , b_j ) $ be an algorithmic network as in Definition~\ref{BdefN_SBB}. 
	%such that $  \mathfrak{N}_{SBB}( N , f , t_{z_0} , 1 , j , s ) $ is well defined.
	Let 
	\[ \myfunc{c \,}{ \mathbb{N} } { \mathfrak{C_{BB}} } { x } { c(x)=y } \]
	be a non-decreasing total computable function such that 
	\[ c(z_0 + f( N, t_{z_0}, 1 ) + 2) \geq c_0 + z_0 + f( N, t_{z_0}, 1 ) + 2 \text{ ,} \] where $ t_{z_0} \in \mathrm{T}(G_t) $.
	Let $ x \in \mathbb{N} $ be an arbitrary number.
	Then, there are constant $ C_7 $ and $ \mathrm{p}_{halt} \in \mathbf{L_U} $ such that, for large enough $ N $ and for every network input $ w \in \mathbf{L_U} $ with $ \lg(N) - C_7 \geq \left| w \right|  $, we will have that the expected local algorithmic synergy of a node $ o_i \in  \mathfrak{P}_{SBB}( N , \mathrm{p}_{halt} ) $ in algorithmic network $  \mathfrak{N}_{SBB}( N , f , t_{z_0} , 1 , j , \mathrm{p}_{halt} ) $ toward solving\footnote{ That is, toward \[ \arraycolsep=0.8pt
		\begin{array}{cccl}
		f_h \; : & \left\{ x \, \middle|  \, \lg(N) - C_7 \geq \left| x \right| \right\} \subset \mathbf{L_U} & \to & \{ h , \overline{ h }\} \subset \mathbf{L_U} \\
		& x & \mapsto & f_h(x) 
		=
		\begin{cases}
		\overline{ h }
		& \textnormal{\small , if $ \mathbf{U} $ does not halt on $ x $ } \\
		h 
		& \textnormal{\small , if $ \mathbf{U} $ halts on $ x $}
		\end{cases}
		\end{array} \] } the halting problem with domain 
	\[ X = \left\{ w \, \middle|  \, \lg(N) - C_7 \geq \left| w \right| \right\} \subset \mathbf{L_U} \] 
	\noindent in $ c' = c(z_0 + f( N, t_{z_0}, 1 ) + 2) $ node cycles is larger than $ x$, i.e.,
	\[
	\lim\limits_{ N \to \infty }
	\mathbf{E}_{  \mathfrak{N}_{SBB} } \left( { \displaystyle{ \myDelta_{ iso }^{ \mathfrak{N}_{SBB}( N , f , t_{z_0} , 1 , j , \mathrm{p}_{halt} ) } } } I_K ( ( o_i , c' , w ) : f_h(w) ) \right) \geq x \text{ ,}
	\]
	with probability arbitrarily close to $1$.

	\begin{proof}
		This proof follows from a combination of Theorem~\ref{BthmHaltingproblem} with Definition~\ref{BdefExpectedlocalsynergySBB} for a sufficiently complex $ h \in \mathbf{L_U} $.
		First, we know from Definition~\ref{BdefExpectedlocalsynergySBB} and Equation~\eqref{BequationSimplifiedlocalsynergy} that
		\begin{equation}\label{BequationBasic}
			\begin{aligned}
				\mathbf{E}_{  \mathfrak{N}_{SBB} } \left( { \displaystyle{ \myDelta_{ iso }^{ \mathfrak{N}_{SBB}( N , f , t_{z_0} , 1 , j , s ) } } } I_K ( ( o_i , c' , w ) : f_h(w) ) \right) 
				= \\
				\frac{ \sum\limits_{ o_i \in  \mathfrak{P}_{SBB}( N , s ) } { \displaystyle{ \myDelta_{ iso }^{ \mathfrak{N}_{SBB}( N , f , t , 1 , j , s )  } } } I_K ( ( o_i , c' , w ) : f_h(w) ) }
				{ N } = \\
				\frac{ \sum\limits_{ o_i \in  \mathfrak{P}_{SBB}( N , s ) }  
				\text{\small $ A ( f_h(w) \, | \,\mathbf{U}(  p_{iso} ( o_i , c' , w ) )  ) -
				A ( f_h(w) \, | \, \mathbf{U}(  p_{net}^{b} ( o_i , c' , w ) ) ) \pm \mathbf{O}(1) $ } }
				{ N }
			\end{aligned} 
		\end{equation}
		Let $ w_{min} $ denote the element of language $ \mathbf{L_U} $ in which its length is the minimum length larger than zero.
		From Definition~\ref{BdefP_SBB}, we know every node belongs to $ \mathbf{L_U} $.
		Therefore, since there always are randomly generated nodes that ignore any input and keep returning $ w_{min} $ as output in any node cycle when running isolated, then, from Definitions~\ref{BdefConcatenation}, \ref{BdefAlgComp}, and~\ref{BdefP_SBB} and the law of large numbers, there is $ \epsilon > 0 $ such that
		\begin{equation}\label{BequationLowerboundisolated}
		\begin{aligned}
			\frac{ \sum\limits_{ o_i \in  \mathfrak{P}_{SBB}( N , s ) }  
				\text{\small $ A ( f_h(w) \, | \,\mathbf{U}(  p_{iso} ( o_i , c' , w ) )  ) $ } }
			{ N } &\geq \\
			\frac{ \sum\limits_{
					 o_i \in \left\{ o \, \middle| \text{\tiny $ \arraycolsep=0.6pt 
					\begin{array}{c}
						 %& f_o(o'_i) = y \\ 
						%\land 
						\mathbf{U}(  p_{iso} ( o , c' , w ) ) = w_{min}
					\end{array} $} \right\} 
					}  
				\text{\small $ A ( f_h(w) \, | \,w_{min} ) $ } }
			{ N } 
			+ \\
			\frac{ \sum\limits_{
					o_i \in \left\{ o \, \middle| \text{\tiny $ \arraycolsep=0.6pt 
						\begin{array}{c}
						%& f_o(o'_i) = y \\ 
						%\land 
						\mathbf{U}(  p_{iso} ( o , c' , w ) ) \neq w_{min}
						\end{array} $} \right\} 
				}  
				\text{\small $ A ( f_h(w) \, | \,	\mathbf{U}(  p_{iso} ( o , c' , w ) ) ) $ } }
			{ N } 
			& \geq
			\epsilon \, A ( f_h(w) \, | \,w_{min} )  + 0 
		\end{aligned}
		\end{equation}
		Now, choose $ \mathrm{p}_{halt} $ with $ \overline{ h }= \left( `` \textit{non-halting} " \right)_{ \mathbf{L_U} } $ and $ h = \left( `` \textit{halting} " \right)_{ \mathbf{L_U} } $ as in Definition~\ref{BdefFunctionhalt}, where\footnote{ The number $ 2 $ is not really necessary here. We chose to employ it in order to avoid minor ambiguities in the asymptotic dominance.}
		\begin{equation}\label{BequationConditionalhmin}
			\begin{aligned}
				A(  h \, | \, w_{min}  ) &> \epsilon^{-1} \, \left( x + 2 \, \mathbf{O}(1) \right) \\
				A(  \overline{ h } \, | \, w_{min}  ) &> \epsilon^{-1} \, \left( x + 2 \, \mathbf{O}(1) \right) \\
				h &\neq \overline{ h }
			\end{aligned}
		\end{equation}
		Furthermore, from Theorem~\ref{BthmHaltingproblem}, there is a constant $ C_7 $ such that, for large enough $ N $ and for every network input $ w \in \mathbf{L_U} $ with $ \lg(N) - C_7 \geq \left| w \right|  $, we have that, for every node $ o_i \in  \mathfrak{P}_{SBB}( N , \mathrm{p}_{halt} ) $, 
		\begin{equation}	
			A ( f_h(w) \, | \, \mathbf{U}(  p_{net}^{b} ( o_i , c' , w ) ) ) = A ( f_h(w) \, | \, f_h(w) ) = \mathbf{O}(1)
		\end{equation} 
		in  $ c(z_0 + f( N, t_{z_0}, 1 ) + 2) $ node cycles with probability arbitrarily close to $1$.
		Therefore, from Equations~\eqref{BequationBasic}, \eqref{BequationLowerboundisolated} and~\eqref{BequationConditionalhmin}, we will have that there is a constant $ C_7 $ such that, for every network input $ w \in \mathbf{L_U} $ with $ \lg(N) - C_7 \geq \left| w \right|  $,
		\begin{equation*}
			\begin{aligned}
				\lim\limits_{ N \to \infty }
				\mathbf{E}_{  \mathfrak{N}_{SBB} } \left( { \displaystyle{ \myDelta_{ iso }^{ \mathfrak{N}_{SBB}( N , f , t_{z_0} , 1 , j , \mathrm{p}_{halt} ) } } } I_K ( ( o_i , c' , w ) : f_h(w) ) \right) \geq \\
				\lim\limits_{ N \to \infty }
				\frac{ \sum\limits_{ o_i \in  \mathfrak{P}_{SBB}( N , \mathrm{p}_{halt} ) }  
					\text{\small $ A ( f_h(w) \, | \,\mathbf{U}(  p_{iso} ( o_i , c' , w ) )  ) -
						\mathbf{O}(1) \pm \mathbf{O}(1) $ } }
				{ N } \geq \\
				\epsilon \epsilon^{-1} \left( x + 2 \, \mathbf{O}(1) \right) - 2 \, \mathbf{O}(1)
			\end{aligned}
		\end{equation*}
		\noindent holds with probability arbitrarily close to $1$.
	\end{proof}
\end{thmundersection}

Indeed, for some universal programming laguages and classical labelings on halting computation and non-halting computation, e.g., $1$ and $0$, respectively, the expected local algorithmic synergy of a node may not be positive.
What Theorem~\ref{BthmLocalsynergy} assures is that, for any chosen universal self-delimited programming language and any arbitrarily chosen $ x $, there are $ h $ and $ \overline{ h } $ that can univocally represent the halting case and the non-halting case, respectively, and that the expected local algorithmic synergy of a node becomes larger than $ x $.

\section{Conclusion and future work}\label{sectionConclusion}

We have studied a particular model of algorithmic networks $ \mathfrak{N}_{SBB} $.
These are composed of randomly generated self-delimiting programs as nodes, which share information accordingly to the synergistic imitation-of-the-fittest protocol (SIFP).
From this model, we studied how to make the nodes asymptotically solve the halting problem as the population grows indefinitely. 
In this way, we have shown how a fixed global information-sharing (or communication) protocol can exploit the power of random generation of individuals and the power of selection made by an irreducibly more powerful environment in order to solve an uncomputable problem.

To this end, we have modified the model introduced in \cite{Abrahao2017published} to enable each node to calculate a partial recursive function for the network input and the latest largest integer shared by the neighbors.
Specifically, this modification was made in the imitation-of-the-fittest protocol (IFP) in \cite{Abrahao2017published}.

First, we proved that, if the population/network size is large enough, the network diameter is small compared to the population/network size, and enough communication rounds (i.e., node cycles) are expended (in particular, matching the network diameter), then every node is expected to solve the halting problem for any program with length dominated by a logarithmic order of the population/network size.
In other words, nodes can emergently solve an increasing number of instances of the halting problem as the population grows indefinitely. 
This way, for algorithmic networks $ \mathfrak{N}_{SBB} $, all instances of the halting problem are statistically covered in the limit when the population grows indefinitely.
This result shows that there is at least one fixed algorithm that can be distributedly run on networked randomly generated universal Turing machines, so that the entire algorithmic network $ \mathfrak{N}_{SBB} $ can compute a function in the Turing degree $ \mathbf{0'} $, if the population/network size is large enough.
Therefore, besides computation time and memory, networked randomly generated environment-evaluable nodes (which we may call \textbf{o}-nodes) can be regarded as a third type of computational resource.
Thus, for algorithmic networks $ \mathfrak{N}_{SBB} $, any set with Turing degree $ \mathbf{0'} $  are indeed decidable if enough (but still finite) time, memory, and \textbf{o}-nodes are given.
As already studied for time hierarchies and space hierarchies, we propose as future research the investigation of \textbf{o}-node hierarchies.
Furthermore, we also propose the investigation of resource-bounded versions of our present results, for example, in the case nodes belong to a time complexity class and the environment (i.e., the machine in which each node is being simulated) belongs to sufficiently higher time complexity class.  

Secondly, we introduced two types of node centralities, in particular,
one for distributed processing and one for network diffusion.
From these and from the previous results, we proved that these two centralities can be intrinsically combined to show that, in the previously described conditions, there is one central node that can solve the halting problem in the minimum amount of communication rounds and only if networked.
This result may help understand how node centralities in network science may be related to emergently privileged nodes in distributed processing.

Third, we introduced one type of algorithmic-informational measure of synergy, bridging previously studied concepts in multivariate information theory for stochastic processes to algorithmic information theory and algorithmic networks.
With this respect, the general problem of synergy in networked computable systems can be translated as the problem of measuring the amount of algorithmic information in an arbitrary collection of nodes strictly necessary to calculate a function, but that could not be obtained by any individual isolated node or by any combination of proper subnetworks of the entire network.
Then, narrowing our approach, we defined a measure of average algorithmic synergistic information for individual nodes in the specific case there is a comparison of the totally networked case with the totally isolated case.
We call it local algorithmic synergy.
Further, we showed that, for any chosen universal self-delimited programming language, one can make the algorithmic networks $ \mathfrak{N}_{SBB} $ produce as much expected local algorithmic synergy of a node as one may want.
In this way, we related the emergent algorithmic complexity in \cite{Abrahao2017published} to a new type of emergent property, in the case, synergy.
Thus, showing how systemic properties commonly studied in complex systems science, such as synergy, can be formalized in the context of networked deterministic systems.
Moreover, with respect to synergy and networked computable systems, our results may help unlocking a formalism to find new mathematical phenomena in future work, such as new types of algorithmic measures of synergy for a comparison of the fully networked case with proper subnetworks.
 
The present article follows a general pursuit of an abstract mathematical theory for systemic properties in complex systems
%, especially, living beings. 
(especially, living systems), such as evolution \cite{Chaitin2012,Hernandez-Orozco2018,Hernandez-Orozco2018a,Chaitin2018,Hernandez-Orozco2018}, emergence of complexity~\cite{Abrahao2017published,Abrahao2018publishedAMS,Hernandez-Orozco2018}, and emergence of creativity~\cite{Abrahao2017published,Abrahao2018c,Chaitin2018}.
This way, as synergy and centralities presented in this article, the theory of algorithmic networks goes toward the direction of establishing formal theories for other common systemic properties usually attributed to complex systems.
In this direction, if one assumes the hypothesis that hypercomputation is possible in Nature, our results in this article have shown how Life might have found a way to synergistically harness the power of selection of individuals in sufficiently random population of individuals, even if every living being remains as a computable system.
This way, the present work may also ``open the gate'' for the study of other systemic properties in future work, such as self-organization and autopoiesis, within the context of distributed deterministic systems.

\section*{Acknowledgments}

Authors acknowledge the partial support from CNPq through their individual grants: F. S. Abrahão (313.043/2016-7), K. Wehmuth (312599/2016-1), and A. Ziviani (308.729/2015-3). 
Authors acknowledge the INCT in Data Science – INCT-CiD (CNPq 465.560/2014-8). Authors also acknowledge the partial support from CAPES/STIC-AmSud (18-STIC-07), FAPESP (2015/24493-1), and FAPERJ (E-26/203.046/2017).
We also thank the comments and critiques from Hector Zenil and Mikhail Prokopenko.

\bibliographystyle{amsplain}
\bibliography{2.2.1-CompleteRefs-Felipe.bib}

\begin{bibdiv}
\begin{biblist}

\bib{Abrahao2015}{thesis}{
      author={Abrah{\~{a}}o, Felipe~S.},
       title={{Metabiologia, Subcomputa{\c{c}}{\~{a}}o e
  Hipercomputa{\c{c}}{\~{a}}o: em dire{\c{c}}{\~{a}}o a uma teoria geral de
  evolu{\c{c}}{\~{a}}o de sistemas}},
        type={Ph.D. thesis},
   publisher={Federal University of Rio de Janeiro (UFRJ)},
        date={2015},
}

\bib{Abrahao2016b}{inproceedings}{
      author={Abrah{\~{a}}o, Felipe~S.},
       title={{Emergent algorithmic creativity on networked Turing machines}},
        date={2016},
   booktitle={The 8th international workshop on guided self-organization at the
  fifteenth international conference on the synthesis and simulation of living
  systems ({ALIFE})},
     address={Canc{\'{u}}n},
         url={http://guided-self.org/gso8/program/index.html},
        note={Extended Abstract available at:
  \url{http://guided-self.org/gso8/program/index.html}},
}

\bib{Abrahao2016}{incollection}{
      author={Abrah{\~{a}}o, Felipe~S.},
       title={{The ``paradox'' of computability and a recursive relative
  version of the Busy Beaver function}},
        date={2016},
   booktitle={{Information and Complexity}},
     edition={1},
      editor={Calude, Cristian},
      editor={Burgin, Mark},
   publisher={World Scientific Publishing},
     address={Singapure},
       pages={3\ndash 15},
         url={{https://doi.org/10.1142/9789813109032_0001}},
}

\bib{Abrahao2018c}{inproceedings}{
      author={Abrah{\~{a}}o, Felipe~S.},
      author={Wehmuth, Klaus},
      author={Ziviani, Artur},
       title={{Expected Emergent Algorithmic Creativity and Integration in
  Dynamic Complex Networks}},
   booktitle={Meeting on theory of computation (etc), congress of the brazilian
  computer society (sbc) 2018},
   publisher={Brazilian Computer Society (SBC)},
     address={Natal},
         url={https://doi.org/10.5281/zenodo.1241236},
        note={Available at: \url{https://doi.org/10.5281/zenodo.1241236}},
}

\bib{Abrahao2018publishedAMS}{article}{
      author={Abrah{\~{a}}o, Felipe~S.},
      author={Wehmuth, Klaus},
      author={Ziviani, Artur},
       title={{Emergent Open-Endedness from Contagion of the Fittest}},
        date={2018},
     journal={Complex Systems},
      volume={27},
      number={04},
      eprint={{https://www.complex-systems.com/abstracts/v27_i04_a03/}},
}

\bib{Abrahao2017published}{article}{
      author={Abrah{\~{a}}o, Felipe~S.},
      author={Wehmuth, Klaus},
      author={Ziviani, Artur},
       title={{Algorithmic networks: Central time to trigger expected emergent
  open-endedness}},
        date={2019sep},
        ISSN={03043975},
     journal={Theoretical Computer Science},
      volume={785},
       pages={83\ndash 116},
      eprint={https://doi.org/10.1016/j.tcs.2019.03.008},
}

\bib{Bedau1998}{article}{
      author={Bedau, Mark~A.},
       title={{Four Puzzles About Life}},
        date={1998apr},
        ISSN={1064-5462},
     journal={Artificial Life},
      volume={4},
      number={2},
       pages={125\ndash 140},
      eprint={http://doi.org/10.1162/106454698568486},
         url={http://www.mitpressjournals.org/doi/10.1162/106454698568486},
}

\bib{Bresciani2018}{incollection}{
      author={Bresciani, Ettore},
      author={D'Ottaviano, {\'{I}}tala Maria~Loffredo},
       title={{Basic concepts of systemics}},
        date={2019oct},
   booktitle={Systems, self-organisation and information: an interdisciplinary
  perspective},
      editor={Alfredo, Pereira~Junior},
      editor={Pickering, William~A},
      editor={Gudwin, Ricardo~Ribeiro},
   publisher={Routledge},
     address={Abingdon, Oxon; New York, NY: Routledge, 2019.},
       pages={248},
         url={https://www.taylorfrancis.com/books/9780429465949},
}

\bib{Calude2002}{book}{
      author={Calude, Cristian~S.},
       title={{Information and Randomness: An algorithmic perspective}},
     edition={2},
   publisher={Springer-Verlag},
        date={2002},
        ISBN={3540434666},
}

\bib{Calude2009}{incollection}{
      author={Calude, Cristian~S.},
       title={{Information: The Algorithmic Paradigm}},
        date={2009},
   booktitle={Formal theories of information},
      editor={Sommaruga, Giovanni},
      series={Lecture Notes in Computer Science},
      volume={5363 LNCS},
       pages={79\ndash 94},
         url={http://link.springer.com/10.1007/978-3-642-00659-3{\_}4},
}

\bib{Chaitin2004}{book}{
      author={Chaitin, Gregory},
       title={{Algorithmic Information Theory}},
     edition={3},
   publisher={Cambridge University Press},
        date={2004},
        ISBN={0521616042},
}

\bib{Chaitin2012}{incollection}{
      author={Chaitin, Gregory},
       title={{Life as Evolving Software}},
        date={2012dec},
   booktitle={A computable universe},
      editor={Zenil, Hector},
   publisher={World Scientific Publishing},
     address={Singapure},
       pages={277\ndash 302},
  url={http://books.google.com/books?hl=en{\&}lr={\&}id=s4AR6y1cp5EC{\&}oi=fnd{\&}pg=PA277{\&}dq=Life+as+Evolving+Software{\&}ots=x-JIx4idV6{\&}sig=B{\_}Kfq3ayPTWX3Lk4zWCDMkboUwA
  http://www.worldscientific.com/doi/abs/10.1142/9789814374309{\_}0015},
}

\bib{Chaitin2013}{book}{
      author={Chaitin, Gregory},
       title={{Proving Darwin: making biology mathematical}},
   publisher={Vintage Books},
        date={2013},
        ISBN={978-1-4000-7798-4},
}

\bib{Chaitin2014}{article}{
      author={Chaitin, Gregory},
      author={Chaitin, Virginia M. F.~G.},
      author={Abrah{\~{a}}o, Felipe~S.},
       title={{Metabiolog{\'{i}}a: los or{\'{i}}genes de la creatividad
  biol{\'{o}}gica}},
    language={Spanish},
        date={2014},
        ISSN={0210-136X},
     journal={Investigaci{\'{o}}n y Ciencia},
      volume={448},
       pages={74\ndash 80},
  url={http://www.investigacionyciencia.es/revistas/investigacion-y-ciencia/la-era-de-los-macrodatos-591/metabiologa-los-orgenes-de-la-creatividad-biolgica-11696},
}

\bib{Chaitin2018}{incollection}{
      author={Chaitin, Virginia M. F.~G.},
      author={Chaitin, Gregory~J.},
       title={{A Philosophical Perspective on a Metatheory of Biological
  Evolution}},
        date={2018},
   booktitle={The map and the territory: Exploring the foundations of science,
  thought and reality},
      editor={Wuppuluri, Shyam},
      editor={Doria, Francisco~Antonio},
   publisher={Springer International Publishing},
     address={Cham},
       pages={513\ndash 532},
}

\bib{Cooper2009}{article}{
      author={Cooper, S.~Barry},
       title={{Emergence as a computability-theoretic phenomenon}},
        date={2009oct},
        ISSN={00963003},
     journal={Applied Mathematics and Computation},
      volume={215},
      number={4},
       pages={1351\ndash 1360},
         url={http://linkinghub.elsevier.com/retrieve/pii/S0096300309004159},
}

\bib{Copeland1998}{article}{
      author={Copeland, B.~Jack},
       title={{Turing's O-machines, Searle, Penrose and the brain}},
        date={1998apr},
        ISSN={0003-2638},
     journal={Analysis},
      volume={58},
      number={2},
       pages={128\ndash 138},
  url={https://academic.oup.com/analysis/article-lookup/doi/10.1093/analys/58.2.128},
}

\bib{Copeland2002}{article}{
      author={Copeland, B.~Jack},
       title={{Hypercomputation}},
        date={2002},
        ISSN={09246495},
     journal={Minds and Machines},
      volume={12},
      number={4},
       pages={461\ndash 502},
}

\bib{Costa2015a}{article}{
      author={Costa, Eduardo~Chinelate},
      author={Vieira, Alex~Borges},
      author={Wehmuth, Klaus},
      author={Ziviani, Artur},
      author={da~Silva, Ana Paula~Couto},
       title={{Time Centrality in Dynamic Complex Networks}},
        date={2015},
        ISSN={02195259},
     journal={Advances in Complex Systems},
      volume={18},
      number={07n08},
      eprint={http://arxiv.org/abs/1504.00241
  http://dx.doi.org/10.1142/S021952591550023X},
         url={http://arxiv.org/abs/1504.00241
  http://dx.doi.org/10.1142/S021952591550023X},
}

\bib{Crnkovic2012a}{article}{
      author={Crnkovic, Gordana~Dodig},
      author={Burgin, Mark},
       title={{Unconventional Algorithms: Complementarity of Axiomatics and
  Construction}},
        date={2012},
        ISSN={10994300},
     journal={Entropy},
      volume={14},
      number={11},
       pages={2066\ndash 2080},
}

\bib{DaCosta2006}{article}{
      author={da~Costa, N.C.A.},
      author={Doria, F.A.},
       title={{Some thoughts on hypercomputation}},
        date={2006jul},
        ISSN={00963003},
     journal={Applied Mathematics and Computation},
      volume={178},
      number={1},
       pages={83\ndash 92},
         url={http://linkinghub.elsevier.com/retrieve/pii/S0096300305008362},
}

\bib{DaCosta2009a}{article}{
      author={da~Costa, N.C.A.},
      author={Doria, F.A.},
       title={{How to build a hypercomputer}},
        date={2009oct},
     journal={Applied Mathematics and Computation},
      volume={215},
      number={4},
       pages={1361\ndash 1367},
  url={https://www.sciencedirect.com/science/article/abs/pii/S0096300309004135},
}

\bib{Dingle2018}{article}{
      author={Dingle, Kamaludin},
      author={Camargo, Chico~Q.},
      author={Louis, Ard~A.},
       title={{Input-output maps are strongly biased towards simple outputs}},
        date={2018dec},
        ISSN={20411723},
     journal={Nature Communications},
      volume={9},
      number={1},
       pages={761},
         url={http://www.nature.com/articles/s41467-018-03101-6},
}

\bib{Agazzi2010}{incollection}{
      author={D'Ottaviano, {\'{I}}tala Maria~Loffredo},
       title={{On the theory of quasi-truth}},
        date={2010},
   booktitle={Series special issues of epistemology: Relations between natural
  sciences and human sciences},
      editor={Agazzi, E},
      editor={{Di Bernardi}, J},
   publisher={Tilguer-Genova Publishing Company},
     address={Genova},
       pages={325\ndash 340},
}

\bib{Downey2010}{book}{
      author={Downey, Rodney~G.},
      author={Hirschfeldt, Denis~R.},
       title={{Algorithmic Randomness and Complexity}},
      series={Theory and Applications of Computability},
   publisher={Springer New York},
     address={New York, NY},
        date={2010},
        ISBN={978-0-387-95567-4},
         url={http://link.springer.com/10.1007/978-0-387-68441-3},
}

\bib{Fernandez2013}{incollection}{
      author={Fern{\'{a}}ndez, Nelson},
      author={Maldonado, Carlos},
      author={Gershenson, Carlos},
       title={{Information Measures of Complexity, Emergence,
  Self-organization, Homeostasis, and Autopoiesis}},
        date={2014},
   booktitle={Guided self-organization: Inception},
      editor={Prokopenko, Mikhail},
      volume={9},
   publisher={Springer, Berlin, Heidelberg},
       pages={19\ndash 51},
         url={http://arxiv.org/abs/1304.1842
  http://link.springer.com/10.1007/978-3-642-53734-9{\_}2},
}

\bib{Fouks1999}{article}{
      author={Fouks, Jean~Denis},
       title={{Towards an algorithmic theory of adaptation}},
        date={1999jul},
        ISSN={03043975},
     journal={Theoretical Computer Science},
      volume={223},
      number={1-2},
       pages={121\ndash 142},
         url={http://linkinghub.elsevier.com/retrieve/pii/S0304397597002119},
}

\bib{Griffith2014a}{incollection}{
      author={Griffith, Virgil},
      author={Koch, Christof},
       title={{Quantifying Synergistic Mutual Information}},
        date={2014},
   booktitle={Guided self-organization: Inception},
      editor={Prokopenko, Mikhail},
      series={Emergence, Complexity and Computation},
      volume={9},
   publisher={Springer, Berlin, Heidelberg},
       pages={159\ndash 190},
         url={http://arxiv.org/abs/1205.4265
  http://link.springer.com/10.1007/978-3-642-53734-9{\_}6},
}

\bib{Grunwald2008}{incollection}{
      author={Gr{\"{u}}nwald, Peter~D.},
      author={Vit{\'{a}}nyi, Paul M.~B.},
       title={{Algorithmic Information Theory}},
        date={2008},
   booktitle={Philosophy of information},
      editor={Adriaans, Pieter},
      editor={van Benthem, Johan},
      series={Handbook of the Philosophy of Science},
      volume={8},
   publisher={Elsevier},
       pages={281\ndash 317},
  url={http://linkinghub.elsevier.com/retrieve/pii/B9780444517265500133},
}

\bib{Hernandez-Orozco2018}{article}{
      author={Hern{\'{a}}ndez-Orozco, Santiago},
      author={Hern{\'{a}}ndez-Quiroz, Francisco},
      author={Zenil, Hector},
       title={{Undecidability and Irreducibility Conditions for Open-Ended
  Evolution and Emergence}},
        date={2018feb},
        ISSN={1064-5462},
     journal={Artificial Life},
      volume={24},
      number={1},
       pages={56\ndash 70},
      eprint={http://doi.org/10.1162/ARTL_a_00254},
         url={https://www.mitpressjournals.org/doi/abs/10.1162/ARTL_a_00254},
}

\bib{Hernandez-Orozco2018a}{article}{
      author={Hern{\'{a}}ndez-Orozco, Santiago},
      author={Kiani, Narsis~A.},
      author={Zenil, Hector},
       title={{Algorithmically probable mutations reproduce aspects of
  evolution, such as convergence rate, genetic memory and modularity}},
        date={2018aug},
        ISSN={2054-5703},
     journal={Royal Society Open Science},
      volume={5},
      number={8},
       pages={180399},
      eprint={1709.00268},
         url={https://arxiv.org/abs/1709.00268
  http://rsos.royalsocietypublishing.org/lookup/doi/10.1098/rsos.180399},
}

\bib{Li1997}{book}{
      author={Li, Ming},
      author={Vit{\'{a}}nyi, Paul},
       title={{An Introduction to Kolmogorov Complexity and Its Applications}},
     edition={2},
   publisher={Springer Science {\&} Business Media},
     address={New York},
        date={1997},
        ISBN={0-387-94868-6},
}

\bib{Lizier2018}{article}{
      author={Lizier, Joseph},
      author={Bertschinger, Nils},
      author={Jost, Juergen},
      author={Wibral, Michael},
       title={{Information Decomposition of Target Effects from Multi-Source
  Interactions: Perspectives on Previous, Current and Future Work}},
        date={2018apr},
        ISSN={1099-4300},
     journal={Entropy},
      volume={20},
      number={4},
       pages={307},
      eprint={http://doi.org/10.3390/e20040307},
         url={http://www.mdpi.com/1099-4300/20/4/307},
}

\bib{Longo2012}{article}{
      author={Longo, Giuseppe},
       title={{Incomputability in physics and biology}},
        date={2012oct},
        ISSN={09601295},
     journal={Mathematical Structures in Computer Science},
      volume={22},
      number={5},
       pages={880\ndash 900},
         url={{http://www.journals.cambridge.org/abstract\_S0960129511000569}},
}

\bib{Mikenberg1986}{article}{
      author={Mikenberg, Irene},
      author={da~Costa, Newton C.~A.},
      author={Chuaqui, Rolando},
       title={{Pragmatic truth and approximation to truth}},
        date={1986mar},
        ISSN={0022-4812},
     journal={The Journal of Symbolic Logic},
      volume={51},
      number={01},
       pages={201\ndash 221},
  url={https://www.cambridge.org/core/product/identifier/S0022481200031686/type/journal{\_}article},
}

\bib{Oizumi2014}{article}{
      author={Oizumi, Masafumi},
      author={Albantakis, Larissa},
      author={Tononi, Giulio},
       title={{From the Phenomenology to the Mechanisms of Consciousness:
  Integrated Information Theory 3.0}},
        date={2014may},
        ISSN={15537358},
     journal={PLoS Computational Biology},
      volume={10},
      number={5},
       pages={e1003588},
      eprint={http://doi.org/10.1371/journal.pcbi.1003588},
         url={http://dx.plos.org/10.1371/journal.pcbi.1003588},
}

\bib{Pan2011}{article}{
      author={Pan, Raj~Kumar},
      author={Saram{\"{a}}ki, Jari},
       title={{Path lengths, correlations, and centrality in temporal
  networks}},
        date={2011jul},
        ISSN={1539-3755},
     journal={Physical Review E},
      volume={84},
      number={1},
       pages={016105},
      eprint={http://doi.org/10.1103/PhysRevE.84.016105},
         url={https://link.aps.org/doi/10.1103/PhysRevE.84.016105},
}

\bib{Prokopenko2014}{book}{
      editor={Prokopenko, Mikhail},
       title={{Guided Self-Organization: Inception}},
      series={Emergence, Complexity and Computation},
   publisher={Springer Berlin Heidelberg},
     address={Berlin, Heidelberg},
        date={2014},
      volume={9},
        ISBN={978-3-642-53733-2},
         url={http://link.springer.com/10.1007/978-3-642-53734-9},
}

\bib{Prokopenko2009}{article}{
      author={Prokopenko, Mikhail},
      author={Boschetti, Fabio},
      author={Ryan, Alex~J.},
       title={{An information-theoretic primer on complexity,
  self-organization, and emergence}},
        date={2009sep},
        ISSN={10762787},
     journal={Complexity},
      volume={15},
      number={1},
       pages={11\ndash 28},
      eprint={http://doi.wiley.com/10.1002/cplx.20249},
         url={http://doi.wiley.com/10.1002/cplx.20249},
}

\bib{Prokopenko2019}{article}{
      author={Prokopenko, Mikhail},
      author={Harr{\'{e}}, Michael},
      author={Lizier, Joseph},
      author={Boschetti, Fabio},
      author={Peppas, Pavlos},
      author={Kauffman, Stuart},
       title={{Self-referential basis of undecidable dynamics: From the Liar
  paradox and the halting problem to the edge of chaos}},
        date={2019jan},
        ISSN={15710645},
     journal={Physics of Life Reviews},
      eprint={http://doi.org/10.1016/j.plrev.2018.12.003},
         url={http://arxiv.org/abs/1711.02456
  https://linkinghub.elsevier.com/retrieve/pii/S1571064519300077},
}

\bib{Rogers1987}{book}{
      author={{Rogers Jr.}, Hartley},
       title={{Theory of Recursive Functions and Effective Computability}},
   publisher={MIT Press},
     address={Cambridge, MA, USA},
        date={1987},
        ISBN={0-262-68052-1},
}

\bib{Shimansky2018}{article}{
      author={Shimansky, Yury~P.},
       title={{Trans-algorithmic nature of learning in biological systems}},
        date={2018aug},
        ISSN={0340-1200},
     journal={Biological Cybernetics},
      volume={112},
      number={4},
       pages={357\ndash 368},
         url={http://link.springer.com/10.1007/s00422-018-0757-y},
}

\bib{Siegelmann1995a}{article}{
      author={Siegelmann, H.T.},
      author={Sontag, E.D.},
       title={{On the Computational Power of Neural Nets}},
        date={1995feb},
        ISSN={00220000},
     journal={Journal of Computer and System Sciences},
      volume={50},
      number={1},
       pages={132\ndash 150},
         url={http://linkinghub.elsevier.com/retrieve/pii/S0022000085710136},
}

\bib{Standish2003}{article}{
      author={Standish, Russell~K.},
       title={{Open-ended artificial evolution}},
        date={2003jun},
        ISSN={1469-0268},
     journal={International Journal of Computational Intelligence and
  Applications},
      volume={03},
      number={02},
       pages={167\ndash 175},
  url={http://www.worldscientific.com/doi/abs/10.1142/S1469026803000914},
}

\bib{Syropoulos2008}{book}{
      author={Syropoulos, Apostolos},
       title={{Hypercomputation: Computing beyond the Church-Turing Barrier}},
        date={2008},
}

\bib{Wehmuth2016b}{article}{
      author={Wehmuth, Klaus},
      author={Fleury, {\'{E}}ric},
      author={Ziviani, Artur},
       title={{On MultiAspect graphs}},
        date={2016},
        ISSN={03043975},
     journal={Theoretical Computer Science},
      volume={651},
       pages={50\ndash 61},
      eprint={http://doi.org/10.1016/j.tcs.2016.08.017},
}

\bib{Wehmuth2017}{article}{
      author={Wehmuth, Klaus},
      author={Fleury, {\'{E}}ric},
      author={Ziviani, Artur},
       title={{MultiAspect Graphs: Algebraic Representation and Algorithms}},
        date={2017},
        ISSN={1999-4893},
     journal={Algorithms},
      volume={10},
      number={1},
       pages={1\ndash 36},
      eprint={http://doi.org/10.3390/a10010001},
         url={http://www.mdpi.com/1999-4893/10/1/1},
}

\bib{Wehmuth2018b}{inproceedings}{
      author={Wehmuth, Klaus},
      author={Ziviani, Artur},
       title={{Centralities in High Order Networks}},
    language={English},
   booktitle={{Meeting on Theory of Computation (ETC), Congress of the
  Brazilian Computer Society (SBC) 2018}},
}

\bib{Wehmuth2015a}{inproceedings}{
      author={Wehmuth, Klaus},
      author={Ziviani, Artur},
      author={Fleury, Eric},
       title={{A unifying model for representing time-varying graphs}},
        date={2015oct},
   booktitle={{Proc. of the IEEE Int. Conf. on Data Science and Advanced
  Analytics (DSAA)}},
   publisher={IEEE},
       pages={1\ndash 10},
         url={http://ieeexplore.ieee.org/document/7344810/},
}

\bib{Zenil2006a}{inproceedings}{
      author={Zenil, Hector},
      author={Hernandez-Quiroz, Francisco},
       title={{On the possible Computational Power of the Human Mind}},
        date={2006feb},
   booktitle={Worldviews, science and us},
   publisher={World Scientific Publishing},
       pages={315\ndash 337},
  url={http://arxiv.org/abs/cs/0605065{\%}0Ahttp://dx.doi.org/10.1142/9789812707420{\_}0020},
}

\end{biblist}
\end{bibdiv}

%%%%%%%%%%%%%%%%%%%%%%%%%%%%%%%%%%%%%

\end{document}